\documentclass[prd,showpacs,altaffilletter,twocolumn,
superscriptaddress,floatfix,nofootinbib,preprintnumbers]{revtex4-1} 

\pdfoutput=1
\usepackage{amsfonts,amsmath,graphicx,bm} 
\usepackage{dcolumn}
\usepackage[pdftex,
       colorlinks=true,
       urlcolor=blue,      
       linkcolor=red,
       pdfpagemode=None,
       bookmarksopen=true]{hyperref}
\usepackage{ifpdf}
\usepackage{multirow}
\usepackage[colorlinks=true]{hyperref}
\usepackage{url}

\newcommand{\cambridge}{Department of Applied Mathematics and
Theoretical Physics, University of Cambridge, Cambridge CB3 0WA, UK}

\newcommand{\glasgow}{SUPA, School of Physics and Astronomy,
  University of Glasgow, Glasgow, G12 8QQ, UK}

\newcommand{\mainz}{Institut f{\"u}r Kernphysik, University of Mainz,
  Becherweg 45, 55099 Mainz, Germany}

\newcommand{\fermilab}{Fermi National Accelerator Laboratory, Batavia, Illinois, 60510, USA}

\newcommand{\VEC}[1]{{\bf \bm{#1}}} 

\newcommand{\order}[1]{ {\mathcal{O}(#1)} } 

\def\today{\number\day\space\ifcase\month\or
January\or February\or March\or April\or May\or June\or
July\or August\or September\or October\or November\or December\fi
\space\number\year}
\newcount\mins \newcount\hours
\def\now{\hours=\time \mins=\time
	\divide\hours by60 \multiply\hours by60 \advance\mins by-\hours
	\divide\hours by60 
	\number\hours:\ifnum\mins<10 0\fi\number\mins }

\begin{document}
\title{Improving the Kinetic Couplings in Lattice Non-Relativistic QCD}

\author{Christine~T.~H.~\surname{Davies}} 
\email[]{christine.davies@glasgow.ac.uk}
\affiliation{\glasgow}

\author{Judd \surname{Harrison}}
\email[]{Judd.Harrison@glasgow.ac.uk}
\affiliation{\glasgow}
\affiliation{\cambridge}

\author{Ciaran \surname{Hughes}} 
\email[]{chughes@fnal.gov}
\affiliation{\fermilab}
\affiliation{\cambridge}

\author{Ronald~R.~\surname{Horgan}} 
\affiliation{\cambridge}

\author{Georg~M.~\surname{von Hippel}} 
\affiliation{\mainz}

\author{Matthew \surname{Wingate}}
\affiliation{\cambridge}

\collaboration{HPQCD Collaboration}
\email[]{http://www.physics.gla.ac.uk/HPQCD}

\preprint{FERMILAB-PUB-18-698-T}

\begin{abstract}

  We improve the non-relativistic QCD (NRQCD) action by comparing the dispersion relation to that of the continuum through $\order{p^6}$ in perturbation theory. The one-loop matching coefficients of the $\order{p^4}$ kinetic operators are determined, as well as the scale at which to evaluate $\alpha_s$ in the $V$-scheme for each quantity. We utilise automated lattice
perturbation theory using twisted boundary conditions as an infrared
regulator. The one-loop radiative corrections
to the mass renormalisation, zero-point
energy and overall energy-shift of an NRQCD $b$-quark are also found. We also explore how a Fat$3$-smeared NRQCD action and changes of the stability parameter $n$ affect the coefficients. Finally, we use gluon field ensembles at multiple
lattice spacing values, all of which include $u$, $d$, $s$ and $c$
quark vacuum polarisation, to test how the improvements affect the
non-perturbatively determined $\Upsilon(1S)$ and $\eta_b(1S)$ kinetic masses, and the tuning of the $b$ quark mass.
\end{abstract}

\maketitle


\section{Introduction}
\label{sec:Intro}

The Standard Model (SM) of particle physics has been incredibly successful
at describing experimental data to date \cite{PDG:2016, QWG:2010}. However, in many ways, this success has been a double-edged sword; while SM predictions have overwhelmingly agreed with experimental measurements within errors, this
has left little room for large new-physics effects to be observed. Consequently, to illuminate any new-physics phenomena high-precision tests
of the SM must be performed. In the $b$-quark sector, the LHCb and BELLE II experiments will generate increasingly precise measurements. In response to this, we make the next level of improvement to the HPQCD collaboration's formulation of the NRQCD action \cite{Dowdall:Upsilon} which has been used for a number of state-of-the-art $b$-physics calculations \cite{Daldrop:Dwave,Dowdall:Hl,Dowdall:BMeson,Brian:LeptonicWidth, Hughes:Hindered,Dowdall:Hyperfine,Dowdall:ErratumHF,Hughes:tetra}.

In this study we will include, for the first time, operators in the NRQCD action which reproduce the correct quark dispersion relation to $\order{p^6}$. Then, with different values of the NRQCD stability parameter $n$, we use lattice perturbation theory to compute the kinetic matching coefficients to  $\order{\alpha_s p^4}$. We remove the unphysical tadpole contributions
 \cite{Lepage:Pert} from the lattice action and give perturbative
 results for two different tadpole improvement programs: the first by using a mean-field improvement parameter in Landau gauge $u_0$ \cite{Lepage:Pert}, and the second via Fat$3$ smearing \cite{Orginos:Fat3}. Additionally, we determine the one-loop (bare-to-pole) mass renormalisation and zero-point energy of the $b$-quark. These can be combined to give the one-loop energy shift of the NRQCD heavy quark,  and added to non-perturbatively obtained static masses to give numerical results which, after converting from lattice units to GeV, can be compared to experimental data. Further, for each of these quantities the scale $\mu = q^*$  at which to evaluate the strong coupling constant defined in the V-scheme is determined using the Brodsky-Lepage-Mackenzie (BLM) procedure \cite{Lepage:Pert, Hornbostel:Scale}.

After perturbatively determining the full one-loop radiative corrections to the kinetic couplings, we non-perturbatively determine the $\Upsilon(1S)$ and $\eta_b(1S)$ energies in order to examine how improving the NRQCD action, both with additional $\order{p^6}$ operators and with the $\order{\alpha_s p^4}$ couplings, reduces the effect of lattice artefacts. 

This paper is organised as follows. In Section \ref{sec:NRQCDAction} 
we describe the improved NRQCD action. In Section \ref{sec:Pert} we match the $\order{\alpha_sp^4, p^6}$ NRQCD dispersion relation to the continuum, describe our tadpole improvement procedures
and how the scale at which to evaluate $\alpha_V$ is
found. Section \ref{sec:PertComp} describes the computational setup of the automated lattice
perturbation theory, while Section \ref{sec:PertResults} gives an analysis of
the perturbative results. Section \ref{sec:NP} gives details of the non-perturbative calculation and Section \ref{sec:NPResults} presents the non-perturbative results.  We summarise our findings in Section \ref{sec:Conclusions}.  


\section{$b$-Quarks Using NRQCD}
\label{sec:NRQCDAction}

Information about processes involving heavy quarks can be 
computed on the lattice using correlation functions constructed from combinations of heavy-quark
propagators. Current lattice ensembles have small enough
lattice spacings and large enough volumes so that
accurate relativistic $c$-quark formalisms (e.g., Highly Improved Staggered Quarks (HISQ) \cite{HISQAction}) are available. Since the $b$-quark
has a Compton wavelength of $\mathcal{O}(0.04)$ fm, most current lattice ensembles cannot resolve relativistic $b$-quarks since $am_b>1$ \cite{PhysRevD.93.094510}\footnote{Combining results at multiple lattice spacing values and multiple heavy quark masses with a highly improved relativistic action does allow results to be obtained at the physical $b$ quark mass \cite{PhysRevD.86.074503}.}.  However, it is well known that $b$-quarks are very
nonrelativistic inside their bound states (with $v_{\text{rel}}^2\approx 0.1$ for
low-lying bottomonium states) and thus using a nonrelativistic
effective field theory, which has a formal expansion in $p/m_b =v_{\text{rel}}$ \cite{Lepage:ImprovedNRQCD}, is very appropriate. This effective
field theory is then discretised as lattice NRQCD \cite{Lepage:ImprovedNRQCD}. 

HPQCD's formulation of lattice NRQCD has already proven successful
in producing accurate $b$-physics results in the literature. For example, the NRQCD
formalism that gave a quark dispersion relation correct to $\order{\alpha_s p^4}$ has already been used to study bottomonium $S$, $P$ and $D$ wave mass splittings \cite{Dowdall:Upsilon,Daldrop:Dwave}, $B$ meson mass
splittings \cite{Dowdall:Hl}, $B$ meson decay
constants \cite{Dowdall:BMeson, Hughes:2017spc}, $\Upsilon$ and $\Upsilon'$ leptonic
widths \cite{Brian:LeptonicWidth}. Subsequently, the spin-dependent $\order{v^6}$
 operators were added to that NRQCD action in order to compute hindered M$1$ radiative decays \cite{Hughes:Hindered}, precise bottomonium hyperfine splittings \cite{Dowdall:Hyperfine,Dowdall:ErratumHF}\footnote{Four-quark operators were also used in this study.} and to aid in the search for $bb\bar{b}\bar{b}$-type bound tetraquarks \cite{Hughes:tetra}. 

Given the increasingly important emphasis being put on high-precision
calculations needed to keep pace with measurements from the LHCb and
BELLE II experiments, we take the next steps in improving the lattice NRQCD action to reduce the systematic uncertainties in future theoretical calculations using it. The first part of this improvement is to add the necessary operators to the aforementioned NRQCD action that reproduce the correct $\order{p^6}$ quark dispersion relation at tree level. 

The NRQCD action that gives rise to a $\order{p^6}$ correct quark dispersion relation, including $\order{v^4}$ interaction operators \cite{Dowdall:Upsilon}, produces a heavy-quark propagator which can be found through the evolution equation
\begin{align}
G({\VEC{x}},t+1)  & = e^{-aH}  G({\VEC{x}},t), \nonumber \\
G({\VEC{x}},t_{\text{src}})  & = \phi({\VEC{x}})
\end{align}
where $\phi(x)$ is a source function and
\begin{align}
e^{-aH} & = \left( 1 - \frac{a\delta H|_{t+1}}{2} \right) \left( 1
  - \frac{aH_0|_{t+1}}{2n} \right)^n U_t^{\dagger}(x) \nonumber \\
&\hspace{0.7cm}  \times \left( 1 -  \frac{aH_0|_t}{2n} \right)^n \left( 1 - \frac{a\delta H|_t}{2}
\right), \label{eqn:NRQCDGreensFunction}  \\
aH_0  & = -\frac{\Delta^{(2)}}{2am_b}, \nonumber \\
 a\delta H  & =  a\delta H_{v^4} + a\delta H_{p^6}; \nonumber  
\end{align}
\begin{align}
a\delta H_{v^4} & = -c_1 \frac{(\Delta^{(2)})^2}{8(am_b)^3} 
+c_2\frac{i}{8(am_b)^2}\left( {\VEC{\nabla \cdot \tilde{E}}} -
  {\VEC{\tilde{E} \cdot \nabla  }} \right)  \nonumber \\ &
-c_3\frac{1}{8(am_b)^2} {\VEC{\sigma \cdot \left( \tilde{\nabla}
      \times \tilde{E} - \tilde{E} \times \tilde{\nabla} \right) }}
\nonumber \\ &
-c_4\frac{1}{2am_b} {\VEC{ \sigma \cdot \tilde{B}}} 
+c_5 \frac{\Delta^{(4)}}{24am_b}
-c_6 \frac{(\Delta^{(2)})^2}{16n(am_b)^2}, \nonumber \\
 \delta H_{p^6} & = - \frac{c_{(p^2)^3}}{16(am_b)^5}\left(1 -
   \frac{(am_b)^2}{6n^2}\right)(\Delta^{(2)})^3 \nonumber \\ 
& -\frac{c_{p^6}}{180am_b} \Delta^{(6)} \nonumber \\ 
& + \frac{c_{p^2p^4}}{48(am_b)^3}( \Delta^{(2)} \Delta^{(4)}). \label{eqn:kinp6}
\end{align}
Here, $am_b$ is the bare $b$-quark mass, $\nabla$ is the symmetric 
lattice derivative, with $\tilde{\nabla}$ the improved version, and
$\Delta^{(2)}$, $\Delta^{(4)}$, $\Delta^{(6)}$ are the lattice discretisations of 
$\sum\limits_{i} D_i ^2$, $\sum\limits_{i} D_i^4$ and  $\sum\limits_i D_i^6$ respectively, with our conventions given in Appendix \ref{app:Der}. ${\VEC{\tilde{E}}}$, ${\VEC{\tilde{B}}}$ are the improved chromoelectric and chromomagnetic
fields, details of which can be found in \cite{Dowdall:Upsilon}. Each of these fields, as well as the covariant derivatives, must be tadpole-improved using the same improvement procedure as in the perturbative calculation of the matching coefficients \cite{Lepage:Pert}. This will be discussed further in Sec.~\ref{sec:Tads}. The parameter $n$ is used to prevent instabilities at large momentum from the kinetic energy operator, and needs to satisfy the constraint $p^2<4nam_b$. A choice of $n=4$ was suitable for values of $am_b$ used in previous non-perturbative studies.  We choose to put the $a\delta H_{p^6}$ corrections into $a\delta H$ rather than alter $aH_0$ so that the kinetic operator 
remains unchanged. This formulation is also consistent with previous
HPQCD NRQCD actions, is symmetric with respect to time reversal and has smaller
renormalisations than other formulations \cite{Lepage:ImprovedNRQCD}. The
rotationally-symmetry breaking operators (which vanish as $a\to 0$) with
coefficients $c_{p^6}$ and $c_{p^2p^4}$ in $\delta H_{p^6}$ remove higher-order discretisation effects from using finite-difference derivatives. The operator with coefficient $c_{(p^2)^3}$ correctly
adds the term proportional to $({\VEC{p}}^2)^3$ into the heavy-quark dispersion relation.

The matching coefficients $c_i$ in the above Hamiltonian take into account the high-energy UV modes from QCD processes that are not present in NRQCD. Each $c_i$ can be fixed by matching a particular lattice NRQCD formalism to full continuum QCD. Each $c_i$ can be expanded perturbatively as
\begin{align}
  c_i &= 1 + c_i^{(1)}\alpha_s + \order{\alpha_s^2},
\end{align}
and, after tadpole improvement \cite{Lepage:Pert}, we expect $c_i^{(1)}$ to be $\mathcal{O}(1)$. In Sec.~\ref{sec:Matching}, we will match the on-shell NRQCD dispersion relation to that of the continuum, and determine $c_1^{(1)}$, $c_6^{(1)}$ and $c_5^{(1)}$. Each of these coefficients should exhibit benign behaviour as a function of $am_b$ in the regime where the NRQCD effective field theory is well-behaved. In contrast, the coefficient may diverge as the effective field theory breaks down as $p\sim \pi/a $ gets too large or $am_b$ gets too small. We take tree level values, $c_i = 1$, for the coefficients appearing in $\delta H_{p^6}$.

We call the NRQCD Hamiltonian presented in Eq.~(\ref{eqn:kinp6}) the
$\order{p^6}$ Hamiltonian, while choosing $a\delta H_{p^6} = 0$
produces the $\order{p^4}$ Hamiltonian. When including the one-loop corrections to $c_1$, $c_6$ and $c_5$, we denote the $\order{p^6}$ NRQCD action as being $\order{\alpha_sp^4, p^6}$, while if $\delta H_{p^6}=0$ then the action is $\order{\alpha_sp^4}$. 

\subsection{One-loop Matching to $\order{\alpha_s p^4, p^6}$ }
\label{sec:Matching}

\begin{figure}[t]
  \centering
  \includegraphics[width=0.23\textwidth]{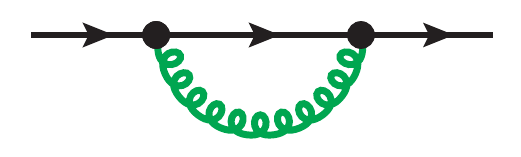}
  \includegraphics[width=0.23\textwidth]{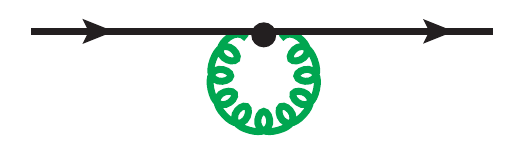}
  \caption{Contributions to the one-loop self energy $\Sigma$ of the heavy quark, showing the {\it{rainbow}} diagram (left) and the {\it{tadpole}} diagram (right). The straight lines represent heavy quarks, while the curly lines represent gluons. }
  \label{fig:SelfEnergy}
\end{figure}

A high-precision non-perturbative calculation of mass splittings will require
knowledge of at least the $\mathcal{O}(\alpha_s)$ corrections to the
matching coefficients in order to improve upon existing few percent errors. For example, when tuning the bare quark mass
$am_b$ fully nonperturbatively in NRQCD, one computes the kinetic mass
of a hadron\footnote{The static mass (the energy corresponding to zero-spatial momentum) in lattice NRQCD  \cite{Dowdall:Upsilon} is shifted due to the removal of the mass term from the Hamiltonian and so one can only
  tune static mass differences fully
  nonperturbatively.} \cite{Dowdall:Upsilon}. This kinetic mass depends on the internal kinematics of the hadron, and hence on (at least) the terms $c_1$, $c_5$, and $c_6$ in the
Hamiltonian. These matching coefficients are known as the kinetic
couplings \cite{Morningstar:KinCouplings}. 

The kinetic couplings can be found perturbatively by matching the
NRQCD on-shell energy (which corresponds to the location of the pole of the quark propagator in the interacting theory) to the continuum 
QCD dispersion relation. From now on, to avoid superfluous notation, we will implicitly work in
lattice units unless otherwise stated. To $\order{\alpha_s}$ the inverse quark propagator may be written
in momentum space as 
\begin{align}
G(p)^{-1} & = G^{(0)}(p)^{-1}- \alpha_s \Sigma(p) \label{eqn:PropInvOne},
\end{align}
with $G^{(0)}(p)^{-1}$ the quark propagator obtained at tree-level
from the non-interacting part of the NRQCD action,
$\Sigma(p)$ the one-loop quark self-energy, $p=(p_4,{\VEC{p}})$ a
four-vector in Euclidean space and
$\omega=-ip_4$ the energy in Minkowski space. The free quark
propagator can be explicitly found as
\begin{align}
  G^{(0)}(p)^{-1} & = \left\{ 1 - e^{-ip_4}F(p)^{2n}F_1(p)^2
  \right\}, \label{eqn:PropInvTree}
\end{align}
\begin{align}
  F(p)   & = 1 - \frac{1}{nm_b} \sum_j \sin^2(p_j/2), \\
F_1(p) & = 1 - \frac{c_5}{3m_b}\sum_j\sin^4(p_j/2) \nonumber \\ & + \frac{\tilde{c}_1}{m_b^3}\left[
  1  + \frac{m_b}{2n} \right] \left[
  \sum_j\sin^2(p_j/2) \right]^2 \nonumber \\
&  - \frac{2c_{(p^2)^3}}{m_b^5}\left[1 - \frac{m_b^2}{6n^2}\right]\left[\sum_j\sin^2(p_j/2)\right]^3 \nonumber \\ 
&  - \frac{8c_{p^6}}{45m_b} \sum_j\sin^6(p_j/2)  \nonumber \\ 
&  + \frac{2c_{p^2p^4}}{3m_b^3} \sum_{j,k} \sin^2(p_j/2)
\sin^4(p_k/2)   \label{eqn:F1p}
\end{align}
where we have defined $\tilde{c}_1 = (c_1 + c_6 m_b/2n)/(1+m_b/2n)$ for
computational ease, and $c_1=c_6=\tilde{c}_1$. The
term $F(p)$ arises from the non-interacting momentum space part of $(1-H_0/2n)$ in 
(\ref{eqn:NRQCDGreensFunction}), while $F_1(p)$ comes from the $(1-\delta H/2)$ piece. 

To find the NRQCD dispersion relation we
determine the on-shell energy $\omega({\VEC{p}})$ which causes a pole
in the full heavy-quark propagator. The one-loop $\omega({\VEC{p}})$
can be found from Eq. (\ref{eqn:PropInvOne}) and
(\ref{eqn:PropInvTree}) as 
\begin{align}
\omega({\VEC{p}}) & =
-\log\left(F^{2n}({\VEC{p}})F_1^2({\VEC{p}})\right) -
\alpha_s\Sigma(\omega_0({\VEC{p}}), {\VEC{p}}) \label{eqn:omega}
\end{align}
 with $\omega_0({\VEC{p}}) = -\log(F^{2n}({\VEC{p}})F_1^2({\VEC{p}}))$, with tree-level coefficients in $F$ and $F_1$, being the tree-level on-shell energy found by setting the tree-level
inverse propagator in Eq. (\ref{eqn:PropInvTree}) to zero. We have constructed the Hamiltonian
in Eq. (\ref{eqn:NRQCDGreensFunction}) to produce a non-relativistic dispersion relation correct to
$\order{p^6}$, and we now include the $\order{\alpha_sp^4}$ correction.
  This yields\footnote{We correct a typographical error in Appendix B of \cite{Dowdall:Upsilon}. }
\begin{align}
\omega_0({\VEC{p}}) & =
  \frac{{\VEC{p}}^2}{2m_b} - \frac{({\VEC{p}}^2)^2}{8m_b^3} +
  \frac{({\VEC{p}}^2)^3}{16m_b^5} \nonumber \\ & + \alpha_s\left\{ c_5^{(1)}
    \frac{{\VEC{p}}^4}{24m_b}  - \tilde{c}_1^{(1)} \left(
      \frac{1}{m_b} + \frac{1}{2n} \right)
    \frac{({\VEC{p}}^2)^2}{8m_b^2}  \right\}. \label{eqn:omega0}
\end{align}
When matching the dispersion relation to  $\order{\alpha_s p^4, p^6}$, it is necessary to
decompose the self-energy $\Sigma(p)$ using the 
small-${\VEC{p}}$ expansion \cite{Morningstar:KinCouplings} as
\begin{align}
\Sigma(p) & = \Sigma_0(\omega) + \Sigma_1(\omega)\frac{{\VEC{p}}^2}{2m_b}
+ \Sigma_2(\omega) \frac{({\VEC{p}}^2)^2}{8m_b^2} +
\Sigma_3(\omega){\VEC{p}}^4 . \label{eqn:Sigma}
\end{align}
Further, when $\omega$ is small, each function has a well-defined
series expansion $\Sigma_m(\omega) =\sum_{l=0}^{\infty}\Sigma_m^{(l)}\omega^l$. The $\Sigma_m^{(l)}$ can be found from derivatives of the quark self-energy as
\begin{align}
& \Sigma_0(\omega)  = \Sigma({{\VEC{p}}={\VEC{0}}}), \\
& \Sigma_1(\omega)  = m_b \left. \frac{\partial^2\Sigma(p)}{\partial p_z^2}\right\vert_{{\VEC{p}}={\VEC{0}}}, \\
& \Sigma_2(\omega)  = m^2_b \left. \frac{\partial^4\Sigma(p)}{\partial
    p_z^2\partial p_y^2}\right\vert_{{\VEC{p}}={\VEC{0}}}, \\
& \Sigma_3(\omega)  =\frac{1}{24} \left. \left( \frac{\partial^4\Sigma(p)}{\partial
    p_z^4} - 3\frac{\partial^4\Sigma(p)}{\partial p_z^2\partial p_y^2} \right) \right\vert_{{\VEC{p}}={\VEC{0}}}, \\
& \Sigma_m^{(l)}  = \left. (-i)^l \frac{1}{l!} \frac{\partial^l
  \Sigma_m(p_4)}{\partial p_4^l}\right\vert_{p_4=0}. 
\end{align}
Then, by using the
tree-level $\omega_0$ from (\ref{eqn:omega0}) in Eq. (\ref{eqn:Sigma}) we find 
\begin{align}
& \Sigma(\omega_0, {\VEC{p}})  = W_0 + \frac{{\VEC{p}}^2}{2m_b}
Z_m^{(1)}  \nonumber \\ 
& \hspace{2.5cm} + \frac{({\VEC{p}}^2)^2}{8m_b^2}\left\{W_1 -
\frac{3Z_m^{(1)}}{m_b}\right\} + W_2{{\VEC{p}}^4}, \label{eqn:SigmaFinal} \\
& m_b^r  = Z_m m_b = m_b\left(1 + \alpha_sZ_m^{(1)} + \order{\alpha_s^2}\right), \\
& Z_m^{(1)}  = \Sigma_0^{(1)} + \Sigma_1^{(0)}, \\
& W_0  = \Sigma_0^{(0)}, \\
& W_1  = 2\Sigma_0^{(2)} + 2\Sigma_1^{(1)} + \Sigma_2^{(0)} +
\frac{2\Sigma_0^{(1)}}{m_b}  + \frac{3\Sigma_1^{(0)}}{m_b}, \label{eqn:W1}\\
& W_2  = \Sigma_3^{(0)} \label{eqn:W2}
\end{align}
where the superscript 'r' denotes renormalised quantities and
$Z_m^{(1)}$ is the $\mathcal{O}(\alpha_s)$ coefficient of the bare-to-pole mass renormalisation. Substituting
(\ref{eqn:F1p}) and (\ref{eqn:SigmaFinal}) into (\ref{eqn:omega}) gives the one-loop NRQCD dispersion relation to $\order{\alpha_s p^4, p^6}$ as
\begin{align}
& \omega({\VEC{p}})  =
  \frac{{\VEC{p}}^2}{2m^r_b} - \frac{({\VEC{p}}^2)^2}{8(m_b^r)^3} +
  \frac{({\VEC{p}}^2)^3}{16(m^r_b)^5}  \nonumber \\
& \hspace{1cm} - \alpha_s \bigg\{  W_0  +
    {\VEC{p}}^4 \left[ W_2 - \frac{ c_5^{(1)}}{24m_b} \right]
    \nonumber \\
& \hspace{2cm} +\frac{({\VEC{p}}^2)^2}{8m_b^2}  \left[ \left(
      \frac{1}{m_b} + \frac{1}{2n} \right)\tilde{c}_1^{(1)} + W_1 \right] \bigg\}. \label{eqn:NRQCDDisp}
\end{align}
Matching Eq. (\ref{eqn:NRQCDDisp}) to the continuum QCD dispersion relation \cite{MovingNRQCD, Dowdall:Upsilon} gives the matching coefficients for $\tilde{c}_1^{(1)},
c_5^{(1)}$ as well as the energy shift of a heavy
quark (to this order) as 
\begin{align}
& \tilde{c}_1^{(1)}  = -\left( \frac{1}{m_b} + \frac{1}{2n}
\right)^{-1} W_1, \label{eqn:c1} \\
& c_5^{(1)}  = 24m_b W_2,  \label{eqn:c5} \\
& C = \omega^{(QCD)} - \omega  = m_b^r  + \alpha_sW_0  = m_b(1 + \alpha_s \delta C), \\
& \delta C  = Z_m^{(1)} + \frac{W_0}{m_b}\,.
\end{align}
The shift $C$ is the perturbative shift of the zero of energy. For each heavy quark in a non-perturbative calculation, the shift can be added to the simulation energy and, after being converted from lattice units to GeV, this can then be compared to
experimental masses \cite{MovingNRQCD,Morningstar:KinCouplings}. In practice hadron masses can be more precisely determined fully non-perturbatively through their kinetic mass in lattice QCD. 

The aim of this study is to determine the one-loop
coefficients $\tilde{c}^{(1)}_1$, $c^{(1)}_5$, $\delta C$ (and
thus also $Z^{(1)}_m$ and $W_0$) for different improved NRQCD actions to find 
the best way forward for increasingly accurate non-perturbative calculations in the future.  Before these coefficients can be used, it is first necessary to remove unphysical contributions from tadpole diagrams which can cause the coefficients to be rather large \cite{Lepage:Pert}.

\subsection{Tadpole Improvement}
\label{sec:Tads}

The authors of Ref.~\cite{Lepage:Pert} show that using
Lie group elements when constructing the lattice field theory introduces
unphysical tadpole diagrams which do not contribute to continuum
schemes. These unphysical tadpole diagrams cause large, process independent renormalisations and produce a poor convergence of the perturbative series. Ref.~\cite{Lepage:Pert} also suggests a solution to this: a gauge-invariant mean-field improvement program (tadpole-improvement) where each lattice link, $U_{\mu}(x)$,
is scaled to $U_{\mu}(x) /u_0$. We choose $u_0$ to be the mean link in Landau gauge, i.e., $u_0 =\langle
\frac{1}{3}\text{Tr} U_{\mu}(x) \rangle $. This mean-field 
parameter has been calculated for the Symanzik-improved gluon
\cite{Dowdall:Upsilon, Hammant:2013} action both perturbatively to one-loop (with $u_0 = 1 - \alpha_su_0^{(2)}$ giving $u_0^{(2)}=0.750$) \cite{Nobes:2001}  and
non-perturbatively \cite{Dowdall:Upsilon, Dowdall:BMeson} (where the
value of $u_0$ depends on the ensemble used, e.g., see Table
\ref{tab:NRQCDParams}). After a tadpole improvement procedure has
been implemented, the one-loop coefficients are expected to be $\order{1}$. The same tadpole-improvement program must of course be implemented in the nonperturbative calculations as has been used for the perturbative calculations. 

Before the mean-field improvement procedure is performed, care must
be taken to ensure that any link-pair cancellations $U^{\dagger}_{\mu}(x)U_{\mu}(x)=1$ occur in the lattice action used in both non-perturbative and perturbative calculations. Such cancellations do not generate any
unphysical tadpole diagrams and scaling by $1/u_0^2$ would be
incorrect. Yet, expanding out the complicated NRQCD Hamiltonian in
(\ref{eqn:NRQCDGreensFunction}) in terms of links $U_{\mu}(x)$ is
excessively expensive for numerical calculations. Consequently,
link-pair cancellations are only taken into account separately for
each derivative, $(\Delta^{(2n)})^m$, or field strengths ($E_i(x)$ or
$B_i(x)$) appearing in the action. This is called partial cancellation \cite{MovingNRQCD, Dowdall:Upsilon}. The difference between the complete and partial cancellation prescriptions was empirically shown not to be sizable \cite{MovingNRQCD}. Formulae for the partially-cancelled derivatives are given in Appendix \ref{app:Der}. 

By using the partially-cancelled mean-field improvement procedure just described, one can find the $\order{\alpha_s}$ tadpole counterterms for the one-loop quantities described in Sec. \ref{sec:Matching}. For the NRQCD action without the $\order{p^6}$ operators, the computation of the tadpole counterterms was checked in two separate calculations. The first was performed analytically, and the second using a Mathematica script. Both calculations reproduced the results of \cite{MovingNRQCD} (in the case where the additional parameter used there, $v$, is set to zero)  and \cite{Dowdall:Upsilon, Gulez:A0V0}. We extended the numerical code to include the $\order{p^6}$ operators. The one-loop tadpole counterterms are given in Appendix \ref{app:Tads}. After the (unimproved) one-loop quantities have been found in lattice perturbation theory, we can add the appropriate tadpole counterterms to determine the improved values. This will be discussed further in Sec.~\ref{sec:PertResults}.

As can be seen in Appendix \ref{app:Tads}, the mean-field counterterms obtained from using the $\order{p^6}$ NRQCD
action contain higher-orders of $1/m_b$ relative to the counterterms
obtained from using the $\order{p^4}$ NRQCD action. This is a
consequence of partially-cancelling the derivative operators
$(\Delta^{(2n)})^m$, whose counterterms are given in Appendix \ref{app:Der}. The impact of this becomes pronounced as $am_b$ is reduced as will become evident in Section \ref{sec:Pert}. 

In this study, we choose to account for the unphysical tadpole contributions
using two different prescriptions. The first prescription proceeds via
the partially-cancelled mean-field improvement procedure just
described. The one-loop tadpole counterterms given in Appendix \ref{app:Tads}, which
depend on $1/am_b$, remove the unphysical tadpoles. As seen
in Sec.~\ref{sec:PertResults}, the improved values give smaller absolute renormalisations compared to the unimproved case, and exhibit a longer plateau over a larger range in $am_b$ indicating stable behaviour in the effective field theory. However, the tadpole counterterms from using the $\order{p^6}$ action diverge faster 
as $am_b\to 0$ due to the higher-order terms in $1/am_b$, and therefore the tadpole-improved one-loop
results obtained from the $\order{p^6}$ NRQCD action also diverge faster (see Sec.~\ref{sec:PertResults}). This could be slightly inconvenient for ensembles with increasingly small lattice spacings, such as the super-fine ensembles currently in use \cite{Fermilab:Vub}, which have a lattice spacing of $a \approx 0.06\mathrm{fm}$. 

Because of this, we explore an
alternative improvement procedure based on the fattening of gauge-links 
\cite{Orginos:Fat3, HIPPY}. The Fat$7$-smeared link \cite{Orginos:Fat7} introduces staples of up to
seven-link paths to completely remove the tree-level couplings to
gluons with high transverse-momentum modes equal to $\pm \pi$. As the
Fat$7$ link is computationally expensive, alternative fat-links have
been designed based on three- or five-link staples, called Fat$3$ and
Fat$5$ respectively. These latter links reduce the couplings to
gluons with high transverse-momentum and suppress unphysical tadpole
diagrams \cite{Orginos:Fat7}. This will be discussed further in section \ref{sec:PertResults}. Therefore, we also explore, for the first time, how a Fat$3$-smeared NRQCD action correct to $\order{p^4}$ affects the renormalisation of kinetic couplings. Here the fattened links are projected back onto $U(3)$ \cite{HIPPY}  (not $SU(3)$). 

The last piece of information needed to use the tadpole-improved one-loop coefficients in a non-perturbative computation is the scale, $q^*$, at which to evaluate the strong coupling constant.

\subsection{Determining the scale of $\alpha_s$}
\label{sec:Scale}

The Brodsky-Lepage-Mackenzie procedure \cite{BLM, Lepage:Pert} determines an optimal $q^*$ for $\alpha_V$, the coupling defined using the heavy quark potential \cite{PhysRevD.28.228}, by examining the momentum flowing through a gluon in the one-loop Feynman diagram. In this prescription, one studies the one-loop integral of a fully dressed gluon within a particular diagram, then uses the running of $\alpha_V(q)$ to find a mean-value $q^*$ which reproduces the integral. To do this, one expands the running of $\alpha_V(q)$ as a polynomial in $\log(q^2/q^{*2})$ and assumes that the leading order log-moments are the dominant contributions. However, in certain areas of parameter space, the leading order log-moments can be anomalously small and give unphysically large or small erroneous $q^*$. This was noticed in \cite{Morningstar:KinCouplings} after which \cite{Hornbostel:Scale} determined $q^*$ when the zeroth and first log-moments are anomalously small via
\begin{align}
\log(q^{*2}) & = \langle\langle \log(q^2) \rangle\rangle \pm [-\sigma^2]^{\frac{1}{2}} \label{eqn:Scale2nd}
\end{align}
where $\langle\langle \log(q^2) \rangle\rangle = \langle
f(q)\log(q^2)\rangle / \langle f(q) \rangle$ indicates the weighted  average, $f(q)$ is the integrand of the one-loop Feynman diagram, and $\sigma^2
= \langle\langle \log^2(q^2) \rangle\rangle - \langle\langle \log(q^2)
\rangle\rangle^2$. The appropriate choice of $\pm$ in
Eq. (\ref{eqn:Scale2nd}) is usually clear based on requiring $q^*$ to
be continuous and physically sensible, although the ambiguity can be
removed by calculating higher log-moments \cite{Hornbostel:Scale}. When
$\sigma^2>0$, only the first term in Eq. (\ref{eqn:Scale2nd}) is
used. However, when $\sigma^2<0$, Eq. (\ref{eqn:Scale2nd}) takes into account the anomalies to first order. 

The unphysical tadpole diagrams contribute to the scale
$q^*$, using the mean-field improvement prescription described in Sec. \ref{sec:Tads} will alter its value. When the tadpole counterterm $c_{tad}\alpha_V(q^*_{tad})$ is added to the one-loop contribution $c_a\alpha_V(q^*_a)$, the second-order formula (\ref{eqn:Scale2nd}) is altered to \cite{Hornbostel:Scale}
\begin{align}
& \log(q^{*2})  = \frac{c_a\langle\langle\log(q^2)\rangle\rangle_a + c_{tad}\langle\langle\log(q^2)\rangle\rangle_{tad}}{c_a + c_{tad}} \pm [-\sigma^2]^{\frac{1}{2}} \label{eqn:ScaleAdd} \\
& \sigma^2 = \frac{c_a\langle\langle\log^2(q^2)\rangle\rangle_a + c_{tad}\langle\langle\log^2(q^2)\rangle\rangle_{tad}}{c_a + c_{tad}} \nonumber \\ 
&  \hspace{1cm} - \left( \frac{ c_a\langle\langle\log(q^2)\rangle\rangle_a + c_{tad}\langle\langle\log(q^2)\rangle\rangle_{tad} }{c_a + c_{tad}}  \right)^2.
\end{align}
Again, if $\sigma^2>0$, then only the first-order term in Eq. (\ref{eqn:ScaleAdd}) is needed and used, while if $\sigma^2<0$  both terms are needed to yield physical results. 

Theoretically, we expect $a\Lambda_{QCD} < aq^* < \pi$ as the one-loop corrections take into account UV modes neglected by imposing a momentum cutoff.  Even though the corrected second order formula given in Eq.~(\ref{eqn:ScaleAdd}) was used, unphysical values of $q^*$ for certain values of $am_b$ in
the one-loop quantities were sometimes obtained. In these cases, although
rare, it was usually clear that the issue was due to the $0^{th}-2^{nd}$
log-moments being anomalously small. To rectify this issue, we use the
simple $n^{th}$-order formula given by \cite{Hornbostel:Scale}
\begin{align}
\log(q^{*2}) & = \frac{\langle f_a(q) \log^n(q^2)\rangle + \langle
  f_{tad}(q) \log^n(q^2)\rangle}{n\langle f_{a}(q) 
  \log^{n-1}(q^2)\rangle + n\langle f_{tad}(q) \log^{n-1}(q^2)\rangle}. \label{eqn:ScaleN}
\end{align}
Leaving the tadpole pieces out of
Eq.~(\ref{eqn:ScaleN}) gives the higher-order tadpole-unimproved scale. As we do not mean-field improve the Fat$3$-smeared one-loop quantities, the above formulae with the tadpole pieces set to zero are used to find $q^*$ in the case of Fat3-smeared links.

\section{Perturbative Determination of One-loop Quantities}
\label{sec:Pert}

\subsection{Perturbative Computational Details}
\label{sec:PertComp}

Due to the complexity of the NRQCD action that we utilise, an efficient computational methodology is needed to calculate the Feynman integrals of the one-loop formulae given in (\ref{eqn:W1}) and (\ref{eqn:W2}). Fortunately, the theory behind the automatic generation of Feynman rules for complex lattice actions exists \cite{Luscher:Pert, HIPPY}. Here, we employ the automated lattice perturbation theory routines HiPPy and HPsrc \cite{HIPPY, HIPPY2}. These routines have been thoroughly tested and used in previous perturbative calculations \cite{Dowdall:Hl, Dowdall:Upsilon, Horgan:Tads, MovingNRQCD, Hammant:2013}. Given that we will produce results for a number of different NRQCD actions, these automated packages are ideal. 

We automatically generate the Feynman rules for a specific NRQCD
action (along with the Symanzik-improved gluonic action \cite{Horgan:Tads, Dowdall:Upsilon}) using the HiPPy package. We can then construct the Feynman diagrams in a generic fashion using the 
HPsrc package, which will use these Feynman rules to numerically evaluate 
the diagram, along with its derivatives thanks to automated 
differentiation techniques \cite{vonHippel:Taylur,HIPPY}.

In the matching procedure both the continuum and lattice contributions
to the dispersion relation are separately infrared (IR) finite. However,
intermediate steps on the lattice may produce IR divergences which
 cancel when evaluating the one-loop quantities. To regulate the IR divergences, we use twisted boundary conditions (TBCs) on a finite-volume lattice where the momentum integral is replaced by
a summation over momentum modes \cite{Luscher:Pert}. TBCs
introduce a lower momentum cutoff by removing the zero mode from the gluon propagator. 
Specifically, we employ triple-twist boundary
conditions with an appropriate squashing factor in the untwisted temporal
direction (used to broaden peaks of the integrand, thus removing most of the 
dependence on $L$ \cite{Luscher:Pert}).  Computational details of both concepts are described in \cite{Horgan:Tads, HIPPY}, and we refer the reader to those articles for
further details. All numerical results are IR finite as
expected. As the dispersion relation is UV finite, this allows us to directly equate results obtained on the lattice to those obtained in the continuum. Furthermore, we test that the gauge-invariant
quantities are independent of the gluon propagator gauge parameter by working in both Feynman gauge and Landau
gauge. All perturbative results
presented, except for the Landau-gauge mean-field parameter $u_0^{(2)}$, will be in Feynman gauge. 

\begin{figure}[t]
  \centering
  \includegraphics[width=0.49\textwidth]{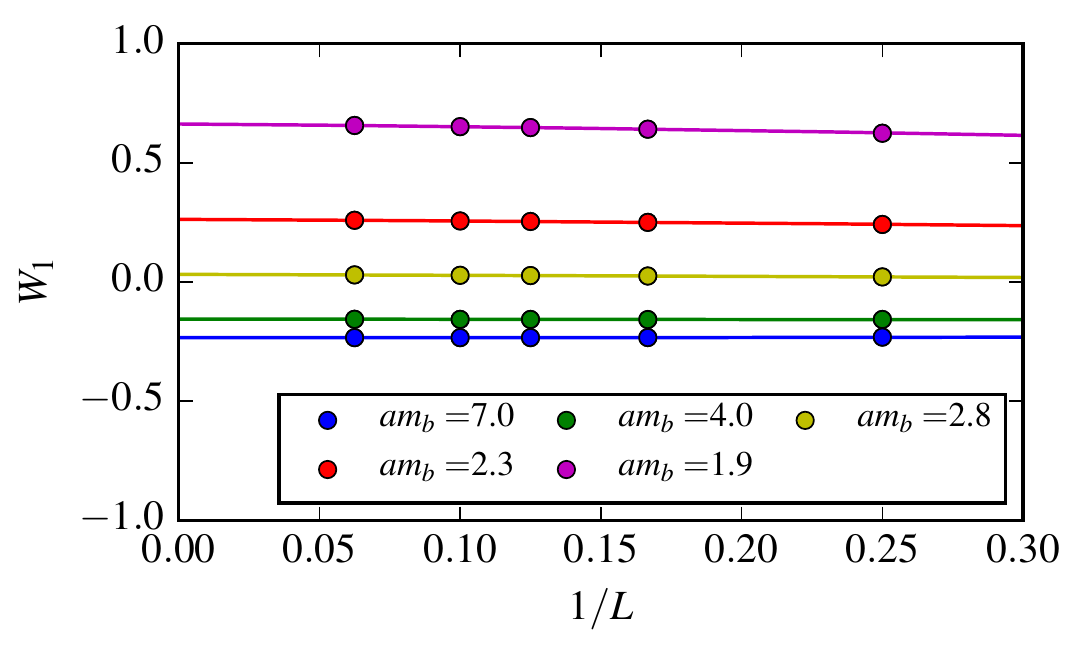}
  \caption{ The raw data for $W_1$ at multiple values of $1/L$ overlaid with our fit curve. }
  \label{fig:W1Extrap}
\end{figure}

The one-loop contributions to the self-energy are shown
in Figure \ref{fig:SelfEnergy}. Care must be taken when numerically evaluating the 
rainbow diagram so that the pole of the heavy-quark propagator does
not cross the temporal integration contour. Details of our implementation of the contour shift can be found in \cite{MovingNRQCD}. 

In this study we will always take the spatial box length to be $4\le
L \le 16$ and choose a temporal extent of $T=16L$. This allows the
pole structure to be resolved in greater detail and reduces finite-$T$  effects. As the one-loop integration is carried out by
direct summation of the twisted momentum modes, numerical results are
exact \cite{Horgan:Tads}. We follow the approach suggested by
\cite{Lepage:Code} in order to fit exact data. Here, our exact
results from TBCs can be expressed as a polynomial in $1/L$
\cite{Luscher:Pert}, yet we are only interested in knowing the
constant term (corresponding to the infinite-volume result). We may
use priors to model the polynomial dependence and then marginalise
\cite{MarjFit} from the exact data the part of the polynomial that
we are not interested in. Using a finite-degree polynomial of order $N$ to model the exact results, 
we find that $N=20$ is a suitable choice and check that all results are unchanged
with its variation. Marginalising the last $N-N_L$ terms of this
polynomial into the exact data and then performing a Bayesian fit
\cite{Lepage:Code, Lepage:Fitting} to a polynomial of degree $N_L$
successfully determines the desired constant parameter of the
polynomial. As is common with marginalised Bayesian fits
\cite{MarjFit}, marginalising all but one or two fit parameters
produces stable and precise results and has seen wide success
\cite{FastFits,Hughes:Hindered}. Even though we  produce successful fits when marginalising all but the constant term, we choose $N_L=5$ and ensure that there is no sensitivity to this. 

\begin{figure}[t]
  \centering
  \includegraphics[width=0.49\textwidth]{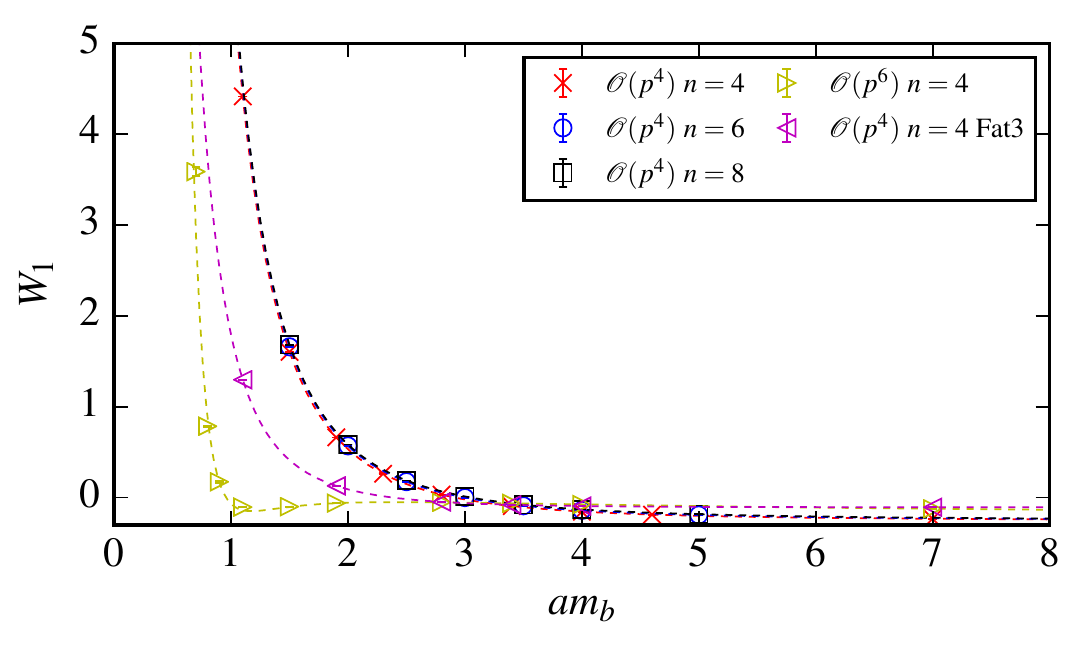}
  \includegraphics[width=0.49\textwidth]{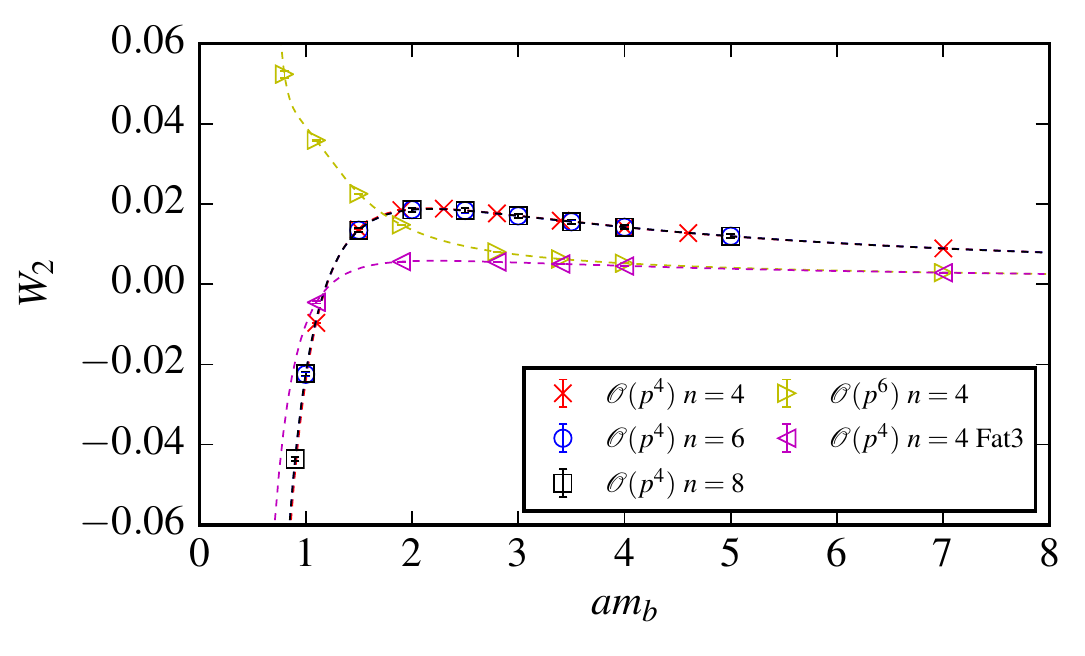}
  \caption{ Numerical values for $W_1$ (top) and $W_2$ (bottom) for
    different NRQCD actions without mean-field improvement as described in
    Sec. \ref{sec:NRQCDAction}. }
  \label{fig:W1W5}
\end{figure}

In the following section, we will give perturbative results for three
NRQCD actions: (i) the $\order{p^6}$ NRQCD action with stability
parameter $n=4$ as described in Sec. \ref{sec:NRQCDAction}; (ii) the
$\order{p^4}$ NRQCD action with stability parameter $n=4,6$ and $8$;
and (iii) a Fat$3$-smeared $\order{p^4}$ NRQCD action with stability parameter $n=4$ and
no mean-field improvement. For a fixed quark and gauge action the
one-loop coefficients depend only on the input parameter $am_b$. We calculate results for a range of $am_b$ 
values, enabling interpolation to values not explicitly calculated 
that may be useful for lattice calculations. This also allows us to demonstrate 
the functional dependence on $am_b$  graphically to see where the divergent behaviour begins as $am_b$ goes to zero. 

\begin{figure}[t]
  \centering
  \includegraphics[width=0.49\textwidth]{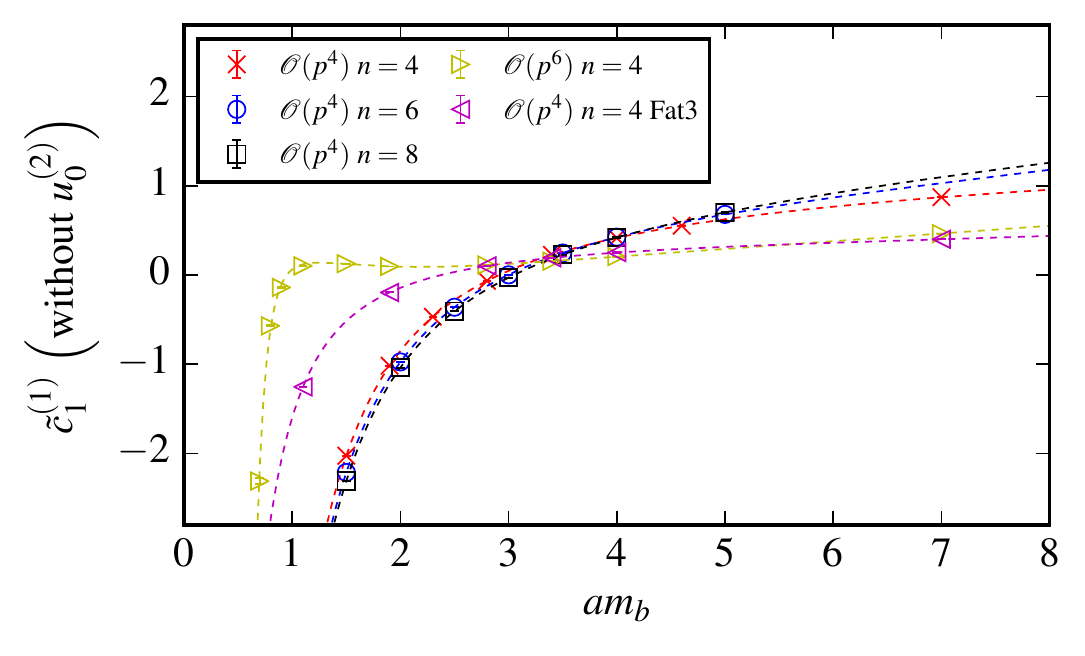}
  \includegraphics[width=0.49\textwidth]{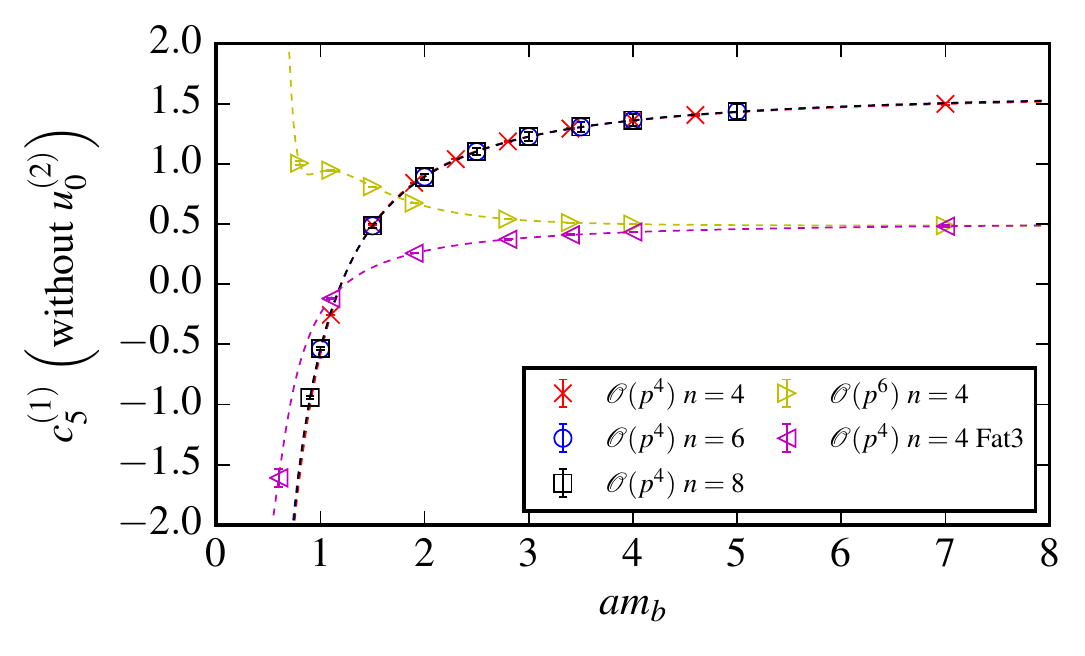}
  \caption{Numerical values for $\tilde{c}_1^{(1)}$ (top) and $c_5^{(1)}$ (bottom) for different NRQCD actions without mean-field improvement. }
  \label{fig:c1cc5c}
\end{figure}

\subsection{Perturbative Results and Analysis}
\label{sec:PertResults}

We calculate $W_1$ and $W_2$ for both the
$\order{p^4}$  and $\order{p^6}$ actions with $n=4$ (including all log moments)
with spatial extent $L=4,6,8,10,16$. We then successfully fit this
data using the methodology described in Sec. \ref{sec:PertComp}. Figure \ref{fig:W1Extrap} shows an example of this, with the raw $W_1$ data at multiple values of $1/L$ overlaid against the fit curve. In fact, we found
 Bayesian fitting to a polynomial so successful that we only needed
data with $L=4,6,8,10$ to obtain the constant term to sub-percent precision in general. Consequently,
we calculate data for $W_1$ and $W_2$ for an $\order{p^4}$ action
with $n=6,8$ and with $L=4,6,8,10$, as well as all $Z_m^{(1)}$,
$W_0$ and Fat3-smeared results (including log moments). The short computational time needed to calculate $u_0^{(2)}$ (and its log
moments), meant that we were able to do this on lattices of size
$L=4,6,8,10,12,14,16$. We present the infinite-volume results for the mean-field unimproved $W_1$ and $W_2$ in Figure \ref{fig:W1W5}. Also shown on each figure is a smooth interpolating curve between the results. This interpolating curve was chosen to be a polynomial in $1/am_b$ in order to reproduce the static limit as $m_b\to \infty$. It is expected that all one-loop quantities diverge as $am_b\to 0$ for our improved NRQCD action, indicating a breakdown of NRQCD, and that is clearly illustrated in our figures.

\begin{figure}[t]
  \centering
  \includegraphics[width=0.49\textwidth]{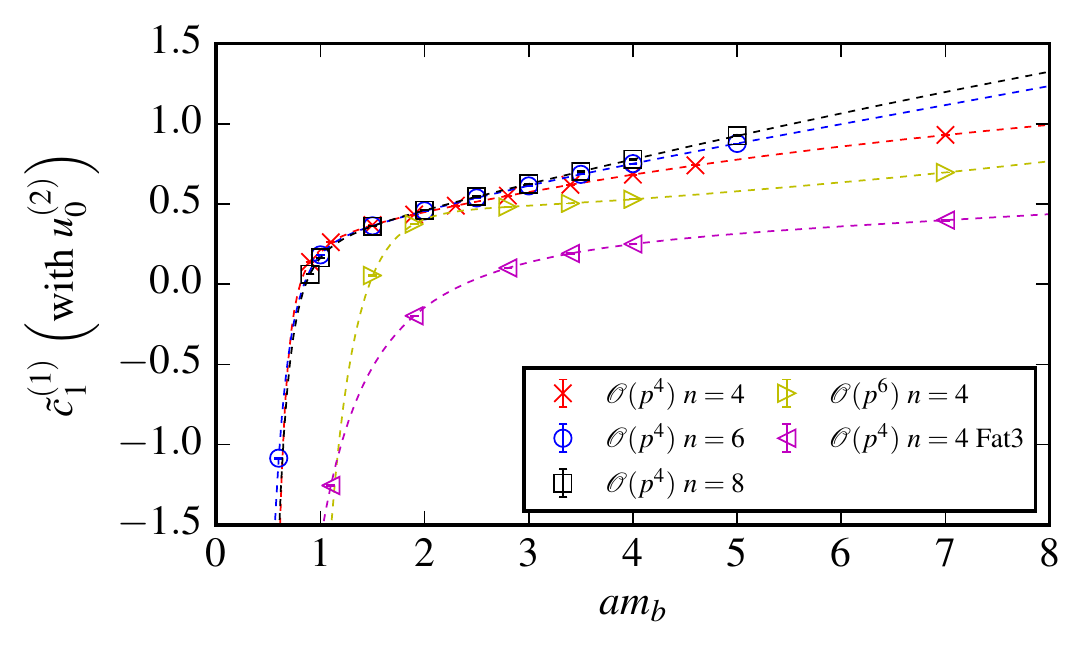}
  \includegraphics[width=0.49\textwidth]{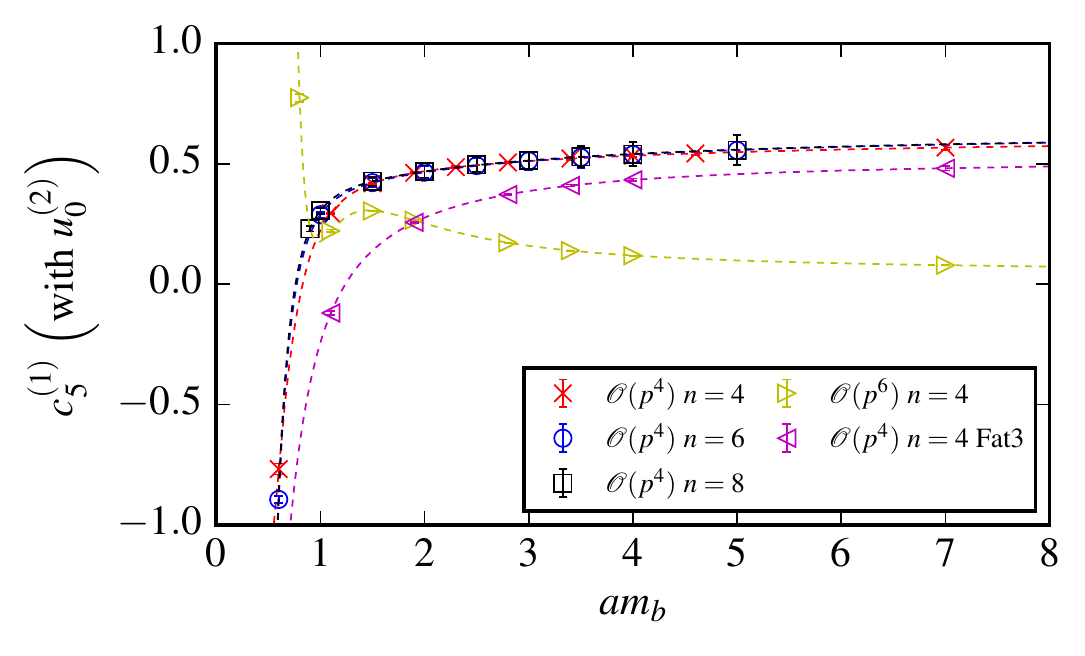}
  \caption{As in Figure \ref{fig:c1cc5c} but with the mean-field improved data
    for $\tilde{c}_1^{(1)}$ (top) and $c_5^{(1)}$
    (bottom). Note the change in vertical scale. The mean-field corrections are given in Appendix
    \ref{app:Tads}. Note that the Fat$3$ smeared data is the same as in
    Figure \ref{fig:c1cc5c}, since no mean-field correction is applied in this case. }
  \label{fig:c1c5}
\end{figure}
\begin{figure}[t]
  \centering
  \includegraphics[width=0.49\textwidth]{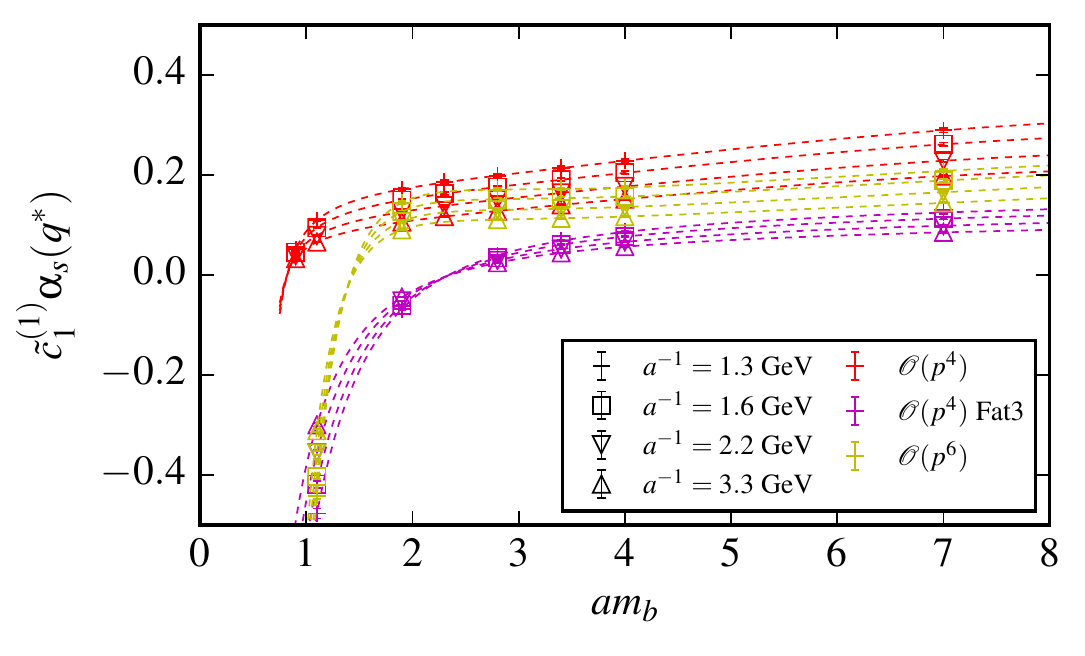}
  \includegraphics[width=0.49\textwidth]{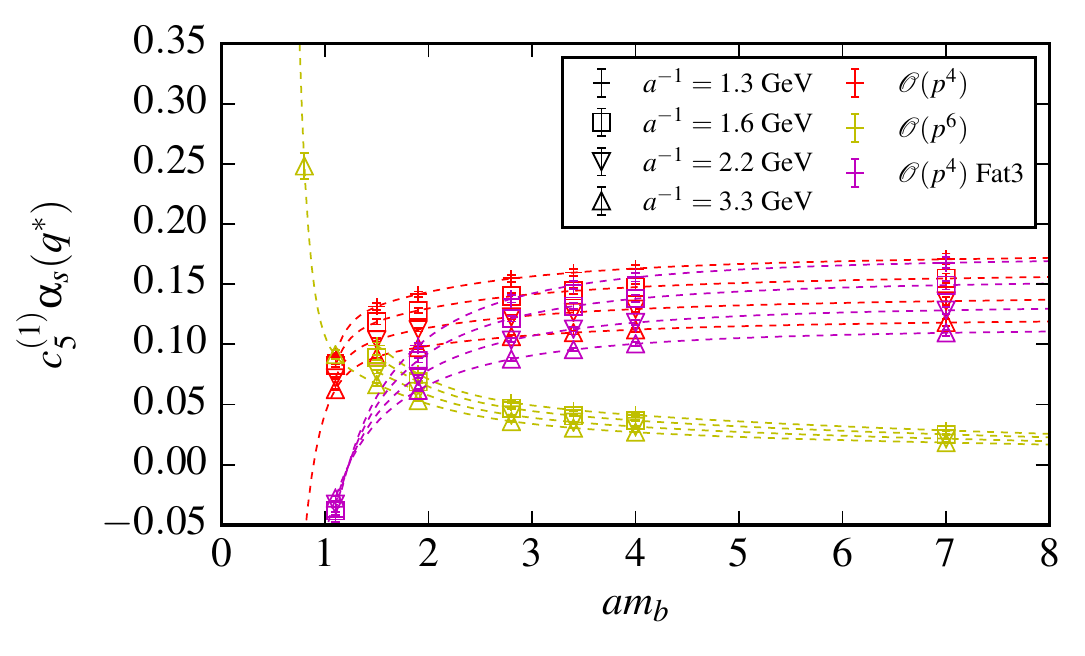}
  \caption{The one-loop radiative correction
    $\tilde{c}_1^{(1)}\alpha_s(q^*)$ (top) and
    $c_5^{(1)}\alpha_s(q^*)$ (bottom) with $\alpha_s$ defined in the $V$-scheme,
  for different NRQCD actions and different values of the lattice
  spacings. We give a subset of the numerical values relevant for
  non-perturbative calculations in Appendix \ref{app:Data}. }
  \label{fig:sc1sc5}
\end{figure}

The difference between $W_1$ and $W_2$ in Figure \ref{fig:W1W5} and
$\tilde{c}^{(1)}_1$ and $c^{(1)}_5$ in Figure \ref{fig:c1cc5c} is purely the conversion 
factors given in Equations (\ref{eqn:c1}) and (\ref{eqn:c5}). In these
plots, a clear observation is that 
the results are very insensitive to an 
increase in $n$ over the mass ranges that we are interested in for non-perturbative calculations on the lattice ($1<am_b<5$). Therefore, as there is no clear benefit to increase $n$ in the perturbative results, future non-perturbative calculations can choose $n=4$ for all $am_b$ in this range. The Fat$3$ smearing works as expected to remove the unphysical tadpoles
 (as outlined in Sec. \ref{sec:Tads}), indicated by a reduction in the absolute size of the
 one-loop corrections. There is a significant improvement, in terms
of longer plateau in $am_b$ and sharper divergence at smaller $am_b$, when using the $\order{p^6}$
 NRQCD action over the $\order{p^4}$. This improved behaviour in the couplings
 leads to the expectation that the second-order couplings are also
 well-behaved. 

In Figure \ref{fig:c1c5}, we then include the mean-field tadpole corrections for all results (except those for Fat$3$ smearing data) with the formulae explicitly given in
Appendix \ref{app:Tads}. The infinite-volume values are given
in Table \ref{tab:Tads}.
Table \ref{tab:Tads} shows why we do not need tadpole-improvement when smeared links are used. The one-loop coefficient in $u_0$ is much smaller in the smeared cases reflecting the fact that tadpole effects are much smaller and the mean smeared link is much closer to $1$. Little is then gained by dividing by it.

As can be seen in Figure \ref{fig:c1c5}, mean-field improvement noticeably reduces
the magnitude of the one-loop coefficients, where it is applied. Due to the higher-order $1/am_b^n$ terms in the
$\order{p^6}$ mean-field counterterms, as described in Sec. \ref{sec:Tads}, the
one-loop couplings with the $\order{p^6}$ action now diverge earlier as
$am_b \to 0$. This is not a desirable feature. This common behaviour is seen in all mean-field improved data we present. Interestingly, the absolute value of $c_5^{(1)}$ is significantly reduced when using a $\order{p^6}$ action. $c_5^{(1)}$ is the coupling which removes the rotational-symmetry breaking operator $\Delta^{(4)}$ at one-loop. Therefore, it is indicative that the $\order{p^6}$ action will reduce $SO(3)$ symmetry breaking in non-perturbative calculations also, as will be discussed in Sec. \ref{sec:NPResults}. 

To fully determine the one-loop shift to the kinetic couplings, the
scale at which to evaluate $\alpha_s$ in the V-scheme needs to be found. We give the mean-field improved $aq^*$ and the Fat$3$ smeared 
$aq^*$ in Appendix \ref{app:Data}. To determine the physical scale,
$q^*$, we use $a^{-1} = 1.3$, $1.6$, $2.2$ and $3.3$ GeV corresponding to 
very coarse, coarse, fine and superfine MILC ensembles used by the
HPQCD collaboration \cite{Dowdall:Upsilon}.  To run the strong coupling in a
particular renormalisation scheme, an 
initial condition needs to be chosen. Here, we use $\alpha_s^{\overline{\text{MS}}}(M_Z, n_f=5)$ taken from the Particle Data Group \cite{PDG:2016}, where  $\alpha_s$ is defined in the $\overline{\text{MS}}$-scheme, $M_Z$ is the $Z$-boson
mass and $n_f=5$ is the number of active flavours. To use this with our data,
we perturbatively remove the $b$-quarks' contribution to the running \cite{AddQuarkSea}, with $m_b(m_b) = 4.164(23)$ GeV \cite{MarjFit}, convert to the $V$-scheme \cite{Schroder:Convert, Lepage:Pert} and run to $q^*$ \cite{QuarkMass:2014}. Finally, we combine $\alpha_{V}(q^*)$ with the one-loop coefficient to give the full one-loop coefficient. These are plotted in
Figure \ref{fig:sc1sc5}. 

\begin{figure}[t]
  \centering
  \includegraphics[width=0.49\textwidth]{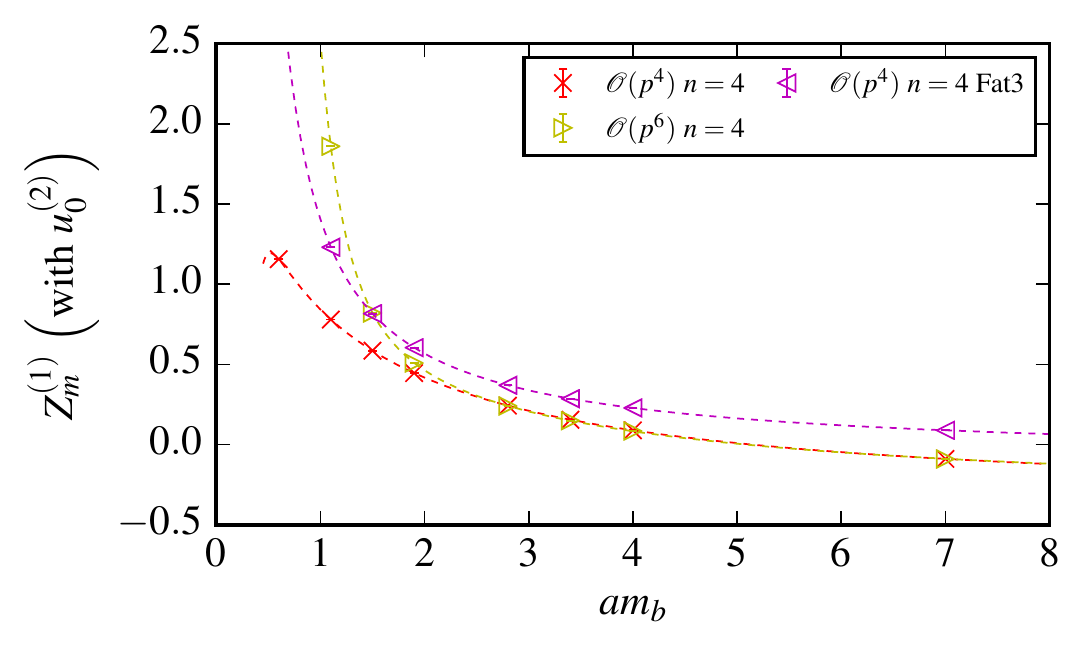}
  \includegraphics[width=0.49\textwidth]{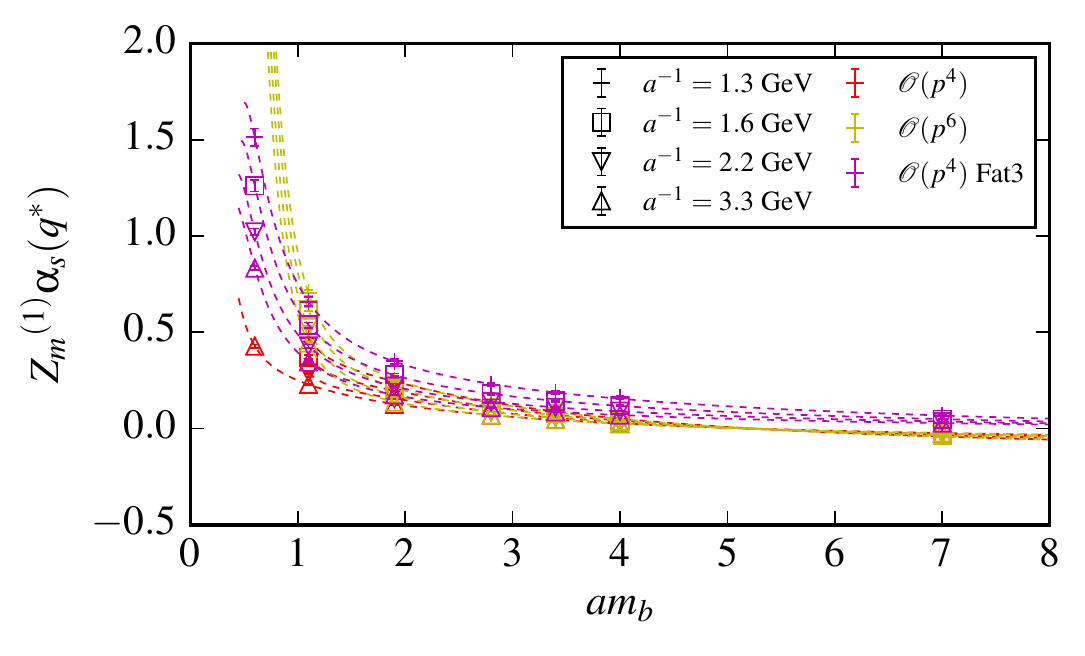}
  \caption{Numerical values for $Z_m^{(1)}$ (top) and the one-loop radiative correction
    $Z_m^{(1)}\alpha_s(q^*)$ (bottom) with $\alpha_s$ defined in the $V$-scheme,
  for different NRQCD actions and different values of the lattice
  spacings. Note that the unsmeared data is mean-field improved as
  described Sec. \ref{sec:Tads}, and the Fat$3$ smeared results are not mean-field improved. }
  \label{fig:zm}
\end{figure}

We show data for the tadpole-improved $Z_m^{(1)}$ in Figure \ref{fig:zm}. Without
mean-field improvement, the $\order{p^4}$ and $\order{p^6}$ data overlap
very closely and are not shown due to this. The modified behaviour after mean-field improvement is
due to the different tadpole-corrections, where
again, the $\order{p^6}$ tadpole-corrections cause faster divergences
as $am_b\to 0$. The mean-field corrections work as expected to reduce the
absolute value of $Z_m^{(1)}$, e.g., $Z_m^{(1)}(am_b=1.9)$ is reduced from $1.57$ (without improvement) to $0.45$. The
full one-loop correction is shown in the lower plot in the same figure.

\begin{figure}[t]
  \centering
  \includegraphics[width=0.49\textwidth]{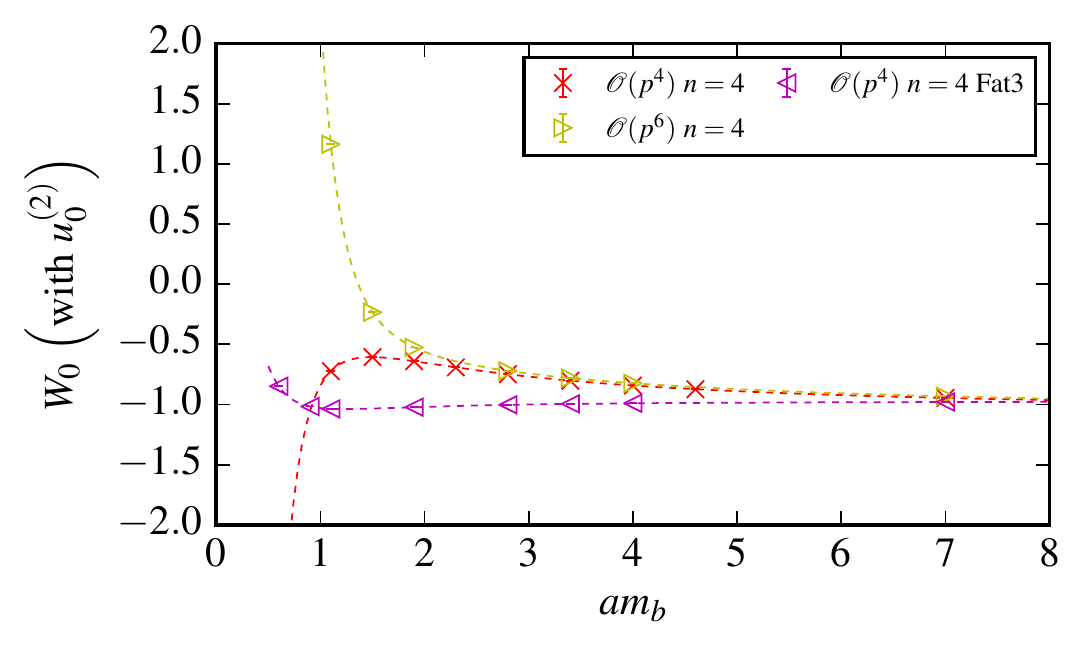}
  \includegraphics[width=0.49\textwidth]{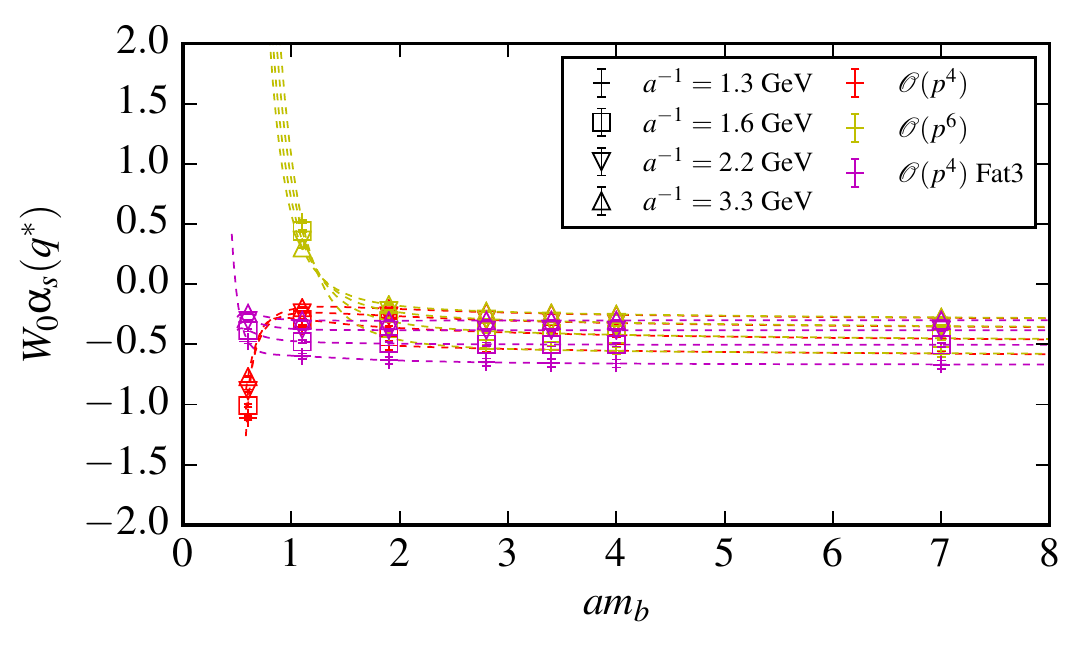}
  \caption{ The same as in Figure \ref{fig:zm} but for $W_0$ in
    place of $Z_m^{(1)}$. }
  \label{fig:E0}
\end{figure}

$W_0$ is
observed to have opposite sign to $Z_m^{(1)}$, but has similar qualitative
features as those just described, e.g., the mean-field unimproved result is typically is a factor of $2 - 3$ in magnitude larger than the mean-field improved
values and there is a clear plateau and a sharp divergence at small $am_b$. The tadpole-improved $W_0$ is shown in Figure \ref{fig:E0}.

\begin{figure}[t]
  \centering
  \includegraphics[width=0.49\textwidth]{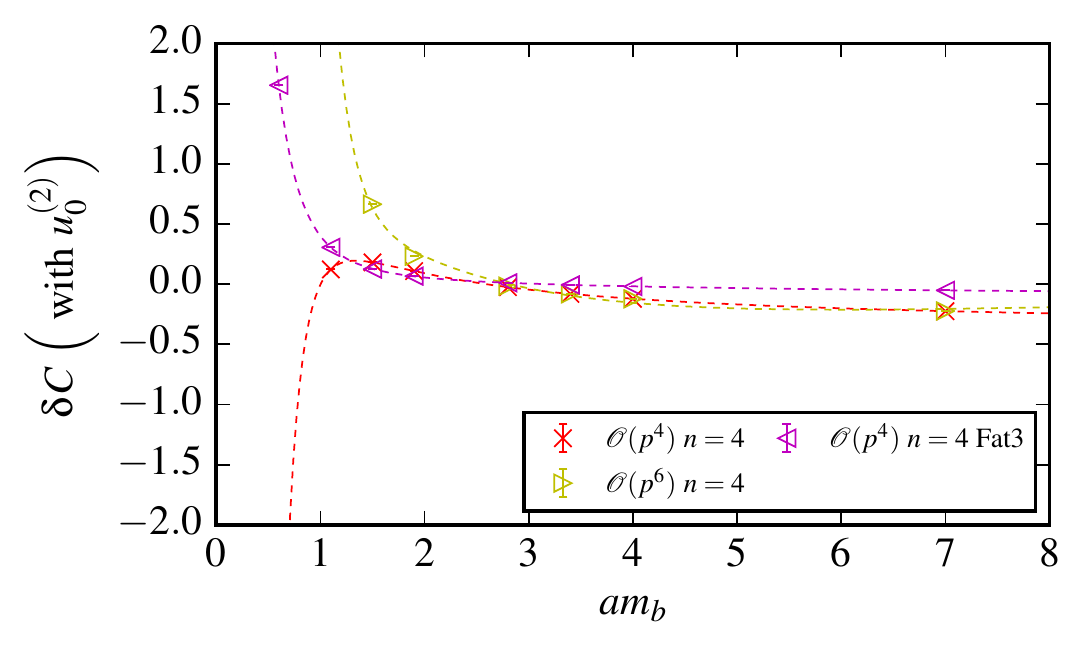}
  \includegraphics[width=0.49\textwidth]{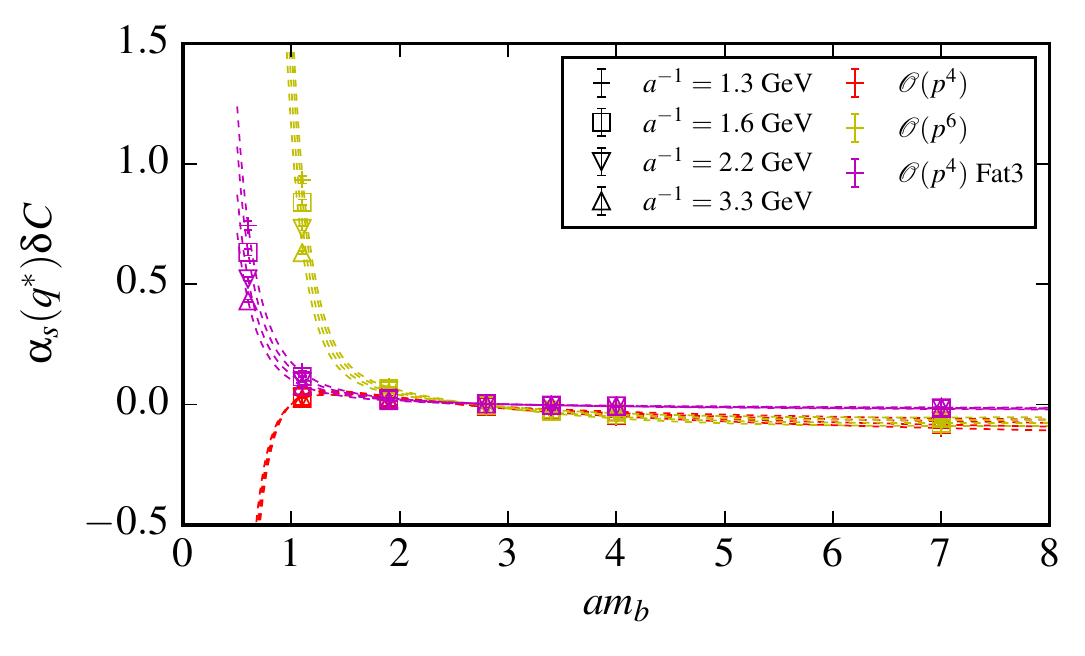}
  \caption{The same as in Figure \ref{fig:zm} but with $\delta C$ in
    place of $Z_m^{(1)}$.}
  \label{fig:shift}
\end{figure}

Because $Z_m^{(1)}$ and $W_0$ have opposite sign,  the one-loop shift in the zero of energy $\delta C$ is found to be very close to zero in all cases as seen in Figure \ref{fig:shift}. 

We note that our $\order{p^4}$  $\tilde{c}^{(1)}_1$, $c_5^{(1)}$ and $Z_m^{(1)}$ differ by small but significant amounts from those in Ref.~\cite{Dowdall:Upsilon}. Ref.~\cite{Dowdall:Upsilon} used Monte Carlo integration combined with numerical derivatives, which they note leads to unstable behaviour when there are large peaks in the IR region. Consequently subtraction functions were used \cite{Eike}. In our study, we avoid these complications by using TBCs as a gauge-invariant IR regulator, and automatic differentiation for the derivatives, which avoids the numerical instabilities arising from finite-differencing schemes \cite{HIPPY, vonHippel:Taylur}.

Finally, the tadpole-improved results for the one-loop coefficients plotted here are given in Appendix
\ref{app:Data}. Subtracting the mean-field corrections (given in
Appendix \ref{app:Tads}) from this data gives the results before tadpole improvement.

\begin{table}[t]
  \caption{ The one-loop mean Landau gauge link \cite{Horgan:Tads} and its log moments for either an unsmeared 
or smeared gauge-link definition.    
    \label{tab:Tads}}
\begin{center}
  \begin{tabular}{l | l l l }
    \hline \hline 
    Gauge-Link & $u_0^{(2)} = \langle f^{tad}\rangle$ & $\langle
    f^{tad}\log(q^2) \rangle$ & $\langle f^{tad}\log^2(q^2)\rangle$ \\ \hline
    Unsmeared & $0.750275(5)$ & $1.45755(2)$ & $3.6022(1)$ \\
    Fat$3$ & $0.231784(5)$ & $0.26101(7)$ & $0.6429(2) $ \\
    Fat$7$ & $0.108244(5)$ & $0.10271(4)$ & $0.4117(2)$ \\
    \hline\hline 
  \end{tabular}
\end{center}
\end{table}

\section{Non-Perturbative Kinetic masses}
\label{sec:NP}

Here we test how improving the NRQCD action as in
Sec. \ref{sec:NRQCDAction} and \ref{sec:Pert} affects the reliability
and accuracy of energies of bottomonium mesons obtained from non-perturbative calculations. 

The static mass (the energy corresponding to zero spatial momentum) in lattice NRQCD is shifted due to the removal of
the mass term from the Hamiltonian \cite{Dowdall:Upsilon}, where we found the one-loop
shift, $C$, in Sec.~\ref{sec:Matching}. Consequently, one can only determine
static mass differences fully nonperturbatively.  However, one can
still obtain kinetic masses \cite{Dowdall:Upsilon, Lewis:Rad1} entirely non-perturbatively via a fully relativistic dispersion relation as 
\begin{align}\label{equationkinmass}
aM_{kin} & = \frac{a^2{\VEC{P}}^2 - a^2\Delta E^2}{2a\Delta E}
\end{align}
where $a\Delta E$ is the energy difference between the meson with
momentum $a{\VEC{P}}$ and the meson at rest. The kinetic mass depends on the internal kinematics of the hadron, and hence on the kinetic terms in the NRQCD action. For example, changing the coefficient of the $ (\Delta^{(2)})^2 /8am_b^3$ term, $c_1$,
from $1$ to $1 + \order{\alpha_s}$ will modify the amount of the
internal kinetic energy that is incorporated into the meson's kinetic
mass, effectively correcting for an $\order{\alpha_s}$ mismatch
between the static and kinetic masses from this operator's contribution
to the binding energy \cite{Dowdall:Upsilon}. The
change would be expected to be $\order{\alpha_sB}$ where
$B$ is the binding energy of $\order{500}$ MeV. This could in
principle be as large as $150-200$ MeV but in practice was found to
be much smaller and around $80$ MeV on coarse and fine lattices (because $c_1^{(1)}$ is small) \cite{Dowdall:Upsilon}. 

Therefore, the kinetic mass
is the ideal candidate on which to test our improvement of the kinetic
part of the action. Furthermore, the kinetic mass is
typically utilised to tune the $b$-quark mass \cite{Dowdall:Upsilon, Lewis:Rad1, Meinel, Dowdall:bquark} and thus
if sizable improvement is seen, this would indicate that improving the kinetic
action would benefit future calculations, where a highly accurate calculation with a
reliable error budget requires knowledge of at least the $\order{\alpha_s}$
corrections to the matching coefficients.\\

In a rotationally invariant theory, the symmetry group is the semi-direct product of the rotational group $SO(3)$ with three translations. The little group of the symmetry group, used to classify energy eigenstates in terms of invariant quantities (e.g., $J^P$ at zero-momentum and
helicity $\lambda$ at non-zero momentum), is broken by a
finite-volume lattice \cite{Moore:2005}. The symmetry of the
lattice discretisation, which breaks $SO(3)$ symmetry at small
distances, does not need to be the same as the symmetry of the finite
volume, which breaks rotational symmetry at
larger distances \cite{Thomas:Helicity}. Here we consider a cubic lattice in a finite
cubic box with PBCs, and so both the lattice
and the boundary break the full $SO(3)$ rotational symmetry of the
continuum to the (double cover) of the octahedral group, $O_h^D$. The
lattice irreducible representations (irreps) for a cubic finite-volume
on a cubic lattice depend on the allowed momenta types \cite{Moore:2005,Thomas:Helicity} (as not all
lattice-momenta are related by an octahedral symmetry)  and
we reproduce them in Table \ref{tab:LittleGroups} for convenience. The energy eigenstates of the lattice Hamiltonian (as obtained from non-perturbative lattice QCD calculations) are classified according to representations of the lattice symmetry group. 

\begin{table}[h]
  \caption{The different little groups relevant for each momentum type in a finite-volume cubic
    lattice with PBC. The momenta are in units of $2\pi/L$ and $n,m,p$
  are non-zero integers with $n\ne m \ne p$ \cite{Moore:2005,
    Thomas:Helicity}. The single cover irreps describe integer spin states. }
\label{tab:LittleGroups}
\begin{center}
  \begin{tabular}{ lll }
    \hline \hline
    Momentum Type ~~ & Little Group   ~~& Irreps  \\
                  & (Double Cover) & (Single Cover) \\
    $(0,0,0)$ & $O_h^D$ & $A_1^{\pm},A_2^{\pm},E^{\pm},T_1^{\pm},T_2^{\pm}$ \\
    $(n,0,0)$ & $Dic_4$ & $A_1,A_2,E_2,B_1,B_2$ \\
    $(n,n,0)$ & $Dic_2$ & $A_1,A_2,B_1,B_2$ \\
    $(n,n,n)$ & $Dic_3$ & $A_1,A_2,E_2$ \\
    $(n,m,0)$ & $C_4$ & $A_1,A_2$ \\
    $(n,n,m)$ & $C_4$ & $A_1,A_2$ \\
    $(n,m,p)$ & $C_2$ & $A$ \\  
    \hline \hline
  \end{tabular}
\end{center}
\end{table}

We denote the energy computed on the lattice for a $\eta_b$ meson with spatial momentum $P$ as $E_{\eta_b}(|aP|)$. Then $E_{\eta_b}(|a{\VEC{P}}|)$ computed with the same $a^2{\VEC{P}}^2$ but with $a{\VEC{P}}$ which lie in different
lattice little groups (e.g., $(3,0,0)$ which has little group $Dic_4$
and $(2,2,1)$ which has little group $C_4$ ) do not need to yield the
same energy within errors. However, as the infinite-volume continuum
limit is taken and full $SO(3)$ symmetry is restored, these energies
should converge. Improving the lattice NRQCD action,
both by adding in higher-order $\order{p^6}$ terms and one-loop radiative
corrections, should reduce $SO(3)$ symmetry breaking and produce the
desired infinite-volume continuum energies more accurately at a given value of the lattice spacing. Examining the non-perturbative energies should indicate this to be the case. 

Improving the NRQCD action will reduce the breaking of $SO(3)$
symmetry due to a cubic lattice. This is because higher-order
rotational-symmetry breaking operators (which vanish as $a\to 0$)
will be increasingly taken into account correctly, e.g., the
$\sum_i\Delta_i^4, \sum_{i,j}\Delta_i^2\Delta_j^4, \sum_i\Delta_i^6$
operators in Eq. (\ref{eqn:NRQCDGreensFunction}). It is perhaps
indicative that including $\order{p^6}$ operators reduces rotational
symmetry breaking, as we found in Sec. \ref{sec:PertResults} that the
one-loop coupling $c_5^{(1)}$, which is constructed in (\ref{eqn:c5})
to remove the rotational-symmetry breaking $\sum_i p_i^4$ terms from the
dispersion relation to one-loop,  gets reduced when improving to the
$\order{p^6}$ NRQCD action. 

In the following we will describe our non-perturbative computational setup as well as discuss how the data from the kinetic masses illustrates the reduction of $SO(3)$ symmetry breaking when improving the kinetic parts of the NRQCD action. 

\subsection{Non-Perturbative Computational Setup}
\label{sec:NonPertSetup}

\begin{table}[t]
  \caption{Details of the gauge ensembles used in this
    study. $\beta$ is the gauge coupling. $a_{\Upsilon}$ is the lattice spacing determined from the $\Upsilon(2S-1S)$ splitting \cite{Dowdall:Upsilon}, where the error combines statistics, experiment and the dominant NRQCD systematic error. $am_q$ are the sea quark masses, $N_s \times N_T$ gives the spatial and temporal extent of the lattices in lattice units and $n_{\text{cfg}}$ is the number of configurations in each ensemble. In column 1 we use the numbering convention for the ensembles from \cite{Dowdall:Upsilon}. Ensemble $1$ is referred to as ``very coarse'',  3 as ``coarse,'' and 5 as  ``fine''. }
\label{tab:GluonEnsembles}
\begin{center}
  \begin{tabular}{l l l l l l l l}
    \hline \hline 
    Set&  $\beta$ & $a_{\Upsilon}$(fm) & $am_l$ &$am_s$ & $am_c$ &
    $N_s \times N_T$ & $n_{\text{cfg}}$  \\ \hline
    $1$ & $5.8$ & $0.1474(15)$ & $0.013$ & $0.065$ & $0.838$ &$16 \times
    48$ & $1020$  \\ \hline
    $3$ & $6.0$ &$0.1219(9)$ & $0.0102$ & $0.0509$ & $0.635$ &$24 \times
    64$ & $1052$ \\\hline
    $5$ & $6.3$ &$0.0884(6)$ & $0.0074$ & $0.037$ & $0.440$ &$32 \times
    96$ & $1008$ \\ \hline \hline
  \end{tabular}
\end{center}
\end{table}

Our computational setup is similar to that in Refs. \cite{Dowdall:Upsilon, Hughes:Hindered} and we point the reader to those texts for specific
details. However, we give a brief overview. We use gauge ensembles generated by the
MILC collaboration \cite{MILC:Configs} with the tadpole-improved
L\"{u}scher-Weisz gauge action \cite{Hart:GluonImprovement} with  $2 + 1 + 1 $ dynamical flavours of HISQ sea quarks \cite{HISQAction}. Details of these ensembles are given in Table \ref{tab:GluonEnsembles}. We use ensembles at three values of the lattice spacing, 
approximately 0.15 fm, 0.12 fm and 0.09 fm, so that we can test 
the changing impact of lattice discretisation effects.

Details of the covariant derivative and chromo-magnetic/electric field
implementation in our NRQCD action can be found in \cite{Dowdall:Upsilon}. Each of these must be
tadpole-improved using the same improvement procedure as in the
perturbative calculation of the matching coefficients in Sec. \ref{sec:PertResults}. We present kinetic masses using the mean-field improvement procedure where, as in
the perturbative results, we take $u_0$ as the mean trace of the gluon field in
Landau gauge, calculated in \cite{Dowdall:Upsilon, Dowdall:BMeson}. The $u_0$ values used for each ensemble are given in Table \ref{tab:NRQCDParams}. We also give in Table \ref{tab:NRQCDParams} the values that we use for the bare $b$ quark mass $am_b$ on each ensemble.

The lattice two-point correlator most naturally encodes information on
meson energies. We use bilinear $b\overline{b}$ interpolating operators, listed in Table \ref{tab:Operators}
with $\Gamma = i\gamma_5, \gamma^j$, which overlap onto
definite $J^{PC} = 0^{-+}, 1^{--}$ energy eigenstates at rest, respectively,
in the infinite-volume continuum version of our theory (which is
rotationally invariant) \cite{Thomas:Helicity}. In \cite{Hughes:Hindered}, as well as \cite{Thomas:Helicity}, it has been shown that at nonzero
momentum, $\mathcal{O}^{\gamma^5}(p)$ is a helicity operator which creates a
definite $\lambda = 0^-$ energy eigenstate, but
$\mathcal{O}^{\gamma^i}(p)$ creates an admixture of $\lambda = 0^+,
\pm 1$, where these $\lambda$ get contributions
from $J^P$ values as listed in the third column of Table \ref{tab:Operators}.
The $\pm$ superscript on the $\lambda=0$ case represents the eigenvalue
$\tilde{\eta} = P(-1)^J$ from the $\hat{\Pi}$ symmetry (a parity
transformation followed by a rotation to bring the momentum direction
back to the original direction) \cite{Thomas:Helicity}. 

Again, following \cite{Dowdall:Upsilon}, we simultaneously fit
multi-exponential functions to the bottomonium meson
correlator at rest and with momentum $a{\VEC{P}}$. Doing so allows the correlations
between the ground states energies to be correctly taken into
account when computing the kinetic mass. We take priors of $0.1(1.0)$ on the
amplitudes, priors on the ground state energies are
estimated from previous results and given a suitably wide width  \cite{Dowdall:Upsilon},
and priors on energy splittings are taken to be $E_{n+1} - E_n= 500(250)$
MeV. To help invert the covariance matrix a singular value decomposition is used with a tolerance of
$10^{-5}$ \cite{ChainedFit, Hughes:Hindered}. We present fit results, following \cite{Dowdall:Upsilon}, for fits including eleven exponentials for Set $1$,  nine exponentials for Set $3$,
and seven exponentials for Set $5$. 
\begin{table}[t]
  \caption{Parameters used for the valence quarks. $am_b$ is the bare
    $b$-quark mass in lattice units, $u_{0L}$ is the tadpole
    parameter \cite{Hughes:Hindered}. $T_p$ is the total propagation time for the $b$-quark
    propagator and $n_t$ is the number of time sources
    used per configuration. The $\order{\alpha_s}$ matching coefficients for $c_1, c_6$ and $c_5$ are taken from Tables \ref{tab:c1shift} and \ref{tab:c5shift}. As explained in Sec. III the $\mathcal{O}(\alpha_s)$ coefficients are functions of $am_b$; the $\alpha_s$ value they are combined with to give $c_1$, $c_5$ and $c_6$ depends on the lattice spacing. The values are different for each version of the NRQCD action tested. As we focus on the improvements made in this study, $c_2,c_3$ and $c_4$ are taken to be their tree-level values of $1.0$.  \label{tab:NRQCDParams}}
\begin{center}
  \begin{tabular}{l l l l l}
    \hline \hline 
    Set&  $am_b$ & $u_{0L}$ & $T_p$ & $n_t$ \\ \hline
    $1$ & $3.40$ & $0.8195$ & $40$ & $16$ \\ \hline
    $3$ & $2.80$ & $0.8349$ & $40$ & $16$ \\ \hline
    $5$ & $1.90$ & $0.8525$ & $48$ & $16$ \\ \hline \hline
  \end{tabular}
\end{center}
\end{table}

\subsection{Non-Perturbative Results and Analysis}
\label{sec:NPResults}

We generate data for $E_{\eta_b}(|a{\VEC{P}}|)$ and
$E_{\Upsilon}(|a{\VEC{P}}|)$ with momenta $a{\VEC{P}} = (0,0,0)$,
$(1,0,0)$, $(1,1,0)$, $(1,1,1)$, $(2,0,0)$, $(2,1,1)$, $(2,2,1)$ and
$(3,0,0)$ in multiples of $2\pi/L$. As discussed above, helicity classifies the energy
eigenstates of the infinite-volume continuum NRQCD theory at non-zero
momentum. Therefore, compared to the zero-momentum case, additional $J^{P}$ states can contribute to the
correlator data at non-zero momentum. The authors of
\cite{Hughes:Hindered} found that when fitting to a $3\times 3$ matrix
of smeared correlators, the first excited state in the fit at non-zero
momentum was the $\chi_{b1}(1P)$, $h_b(1P)$ for the operators
$\mathcal{O}^{\gamma^5}(x)$, $\mathcal{O}^{\gamma^i}(x)$
respectively. At zero momentum, the first excited state was the
$\eta_b(2S)$, $\Upsilon(2S)$ respectively. By using the
same smearing types and correlators as those authors, we check
that the additional states are present at non-zero momentum when using
a $3\times 3$ fit. However even when a fit does not resolve
the additional (first excited) state accurately we find that the ground state is
uncontaminated and still precise.  Further, the finite-volume lattice
breaks $SO(3)$ symmetry and allows mixing with higher $J^P$ states within each of the lattice
irreps given in Table \ref{tab:LittleGroups}. As in
\cite{Hughes:Hindered}, we find no signal for any mixing in the low-lying spectrum. We conclude that our ground state energies are reliably determined. 

Each $E_{\Upsilon}(|a{\VEC{P}}|)$ extracted from our lattice calculation has
larger errors than those on $E_{\eta_b}(|a{\VEC{P}}|)$ because of the slightly poorer 
signal-to-noise ratio. The statistical errors also grow
with momentum. Consequently, $\Delta E(|a{\VEC{P}}|)$ has larger
absolute errors as $a^2{\VEC{P}}^2$ grows, but the relative error on
$\Delta E(|a{\VEC{P}}|)$ is larger for small $a^2{\VEC{P}}^2$. As a result, the kinetic masses with smaller $a^2{\VEC{P}}^2$ have larger errors, which then stabilise.

On each ensemble we examine the kinetic mass, given by Eq.~(\ref{equationkinmass}), in order to see how the kinetic mass changes for a given $am_b$ as a function of momentum, as we improve the NRQCD action. Since the energies and kinetic masses should only depend on 
the magnitude of the spatial momentum, rotational symmetry breaking effects show up most clearly as 
a difference between the energies corresponding to momenta $(3,0,0)$ 
and $(2,2,1)$ in units of $2\pi/L$, and also as a kinetic mass that depends on
$a^2{\VEC{P}}^2$. 

\begin{table}[t]
  \caption{The local bilinear operators used in this study. Note the
    $i\gamma^5$ is needed to make the overlaps real \cite{Dudek:CharmRad}. The second
    column gives the $J^{PC}$ states that these operators create at
    rest in an infinite volume continuum. The
    third column gives the helicity eigenvalues $\lambda$ that these
    operators create at nonzero momentum in an infinite volume
    continuum which is only rotationally invariant, while the $J$ in brackets
    are the states which contribute to that helicity (c.f.
    \cite{Hughes:Hindered, Thomas:Helicity}.) }
\label{tab:Operators}
\begin{center}
  \begin{tabular}{l l l}
    \hline\hline
    $\mathcal{O}^{\Gamma}(x)$ & $J^{PC}$ & \multicolumn{1}{l}{$\lambda (\leftarrow J^P)$}
    \\ \hline

    $\bar{\psi} i\gamma^5 \psi$ & $0^{-+}$ & $0^- (\leftarrow
    J^P = 0^-,1^+,2^-,\ldots)$ \\ \hline
    \multirow{2}{*}{$\bar{\psi} \gamma^i \psi$} &
    \multirow{2}{*}{$1^{--}$ } & $0^+ (\leftarrow J^P =
    0^+,1^-,2^+,\ldots)$ \\ 
    & & $|1| (\leftarrow J = 1,2,3,\ldots )$ \\\hline\hline
  \end{tabular}
\end{center}
\end{table}

One feature of the results is that the kinetic mass for the $\eta_b$ is slightly larger than that of the $\Upsilon$, rather than being lower to reflect the ordering of the masses seen 
in experiment. This was also seen in \cite{Dowdall:Upsilon} and explained there as the 
result of not including relativistic corrections to the $\sigma \cdot B/2m $
term in the NRQCD action (the term with coefficient $c_4$ in 
Eq. (\ref{eqn:kinp6})). Such corrections (spin-dependent terms at $\mathcal{O}(v^6)$) 
would allow the effect of the $\sigma \cdot B$ term to be correctly 
incorporated in the kinetic mass and solve this problem \cite{Dowdall:Hyperfine,Dowdall:ErratumHF}. 
The strategy adopted in \cite{Dowdall:Upsilon} to mitigate this problem was 
to use the spin-averaged kinetic mass, which is less 
sensitive to these effects, to tune the $b$ quark 
mass. The spin-averaged kinetic mass is given by 
\begin{equation}\label{spinaveq}
M_{kin}^\text{spin-averaged} = \frac{3 M_{{kin},\Upsilon} +M_{{kin},\eta_b}}{4}.
\end{equation}

\begin{figure}[t]
  \centering
  \includegraphics[width=0.45\textwidth]{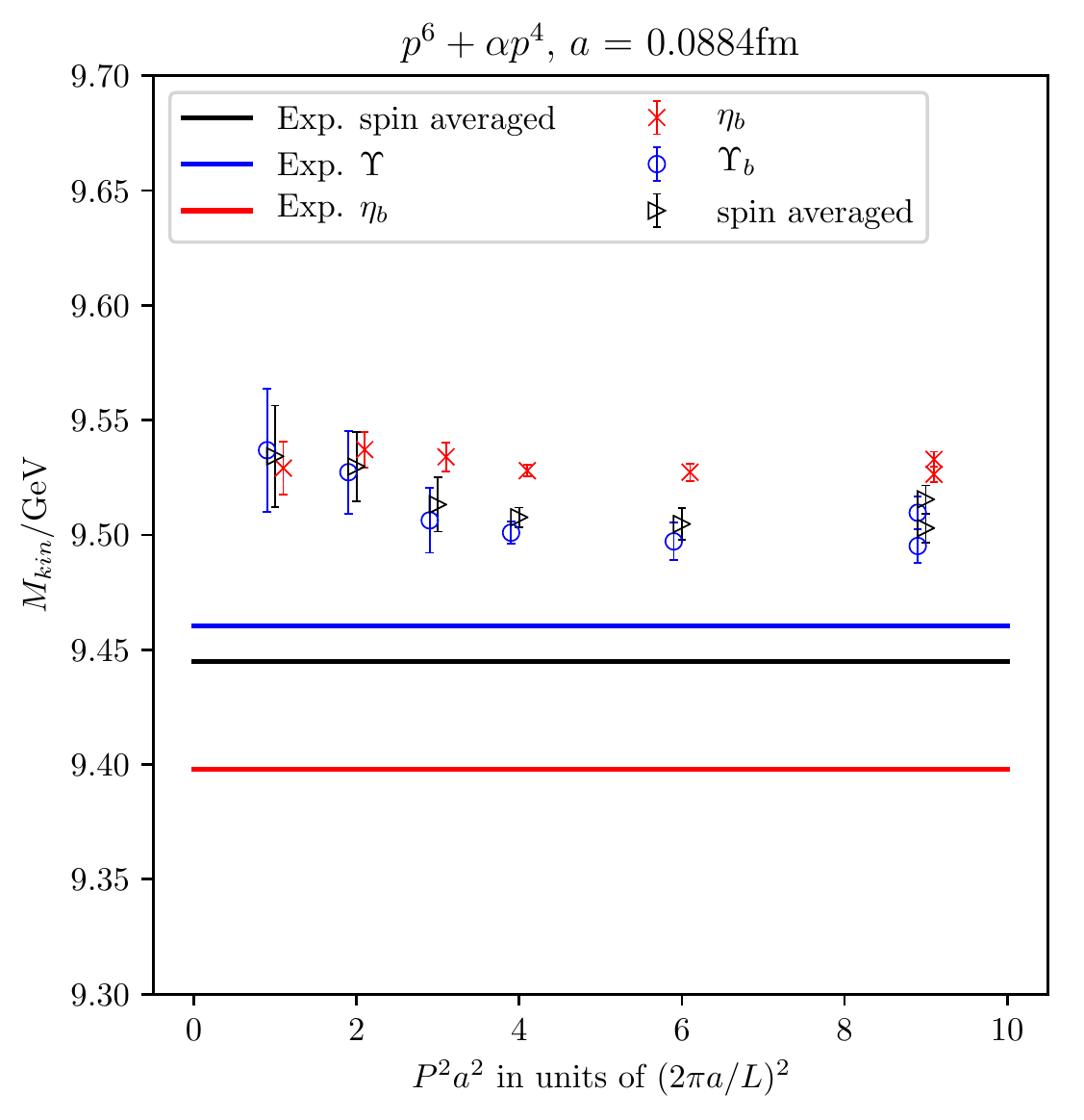}
  \caption{Non-perturbatively obtained values for $M_{kin}$ for the $\eta_b$ and the $\Upsilon$ plotted against momentum squared, together with the spin-averaged kinetic mass, on Set 5. The solid lines show experimental values and errors are statistical only.}
  \label{fig:spinaveragemotivation}
\end{figure}

Figure \ref{fig:spinaveragemotivation} illustrates this feature by 
showing results for the $\Upsilon$ and $\eta_b$ 
kinetic masses on set 5, for the $p^6+\alpha_sp^4$ action. 
We also show the spin-averaged kinetic mass. 
The solid lines show the corresponding experimental 
values. In a full nonperturbative calculation we would 
want to tune the $am_b$ value for each action 
separately to match the spin-averaged kinetic mass 
to experiment. Here however we keep the same $am_b$ 
value for each action on a given ensemble (with only an 
approximate tuning)
so that we can compare how the kinetic 
mass changes.

Data for the spin-averaged kinetic masses from the tree-level
$\order{p^4}$ and $\order{p^6}$ NRQCD actions, both with and without
$\order{\alpha_s p^4}$ corrections, are presented in Figures
\ref{fig:KinE1} and \ref{fig:KinE3} for the ensembles
described in Table \ref{tab:GluonEnsembles}. Errors are statistical
only. Since these plots have the same vertical scale we can see a reduction in the size of $ap^6$ and $\mathcal{O}(\alpha_s)$ effects as the lattice spacing is reduced, from the fact that the range of results become more compressed from Sets 1 to 5. 

In Figure \ref{fig:deltaEetaups} we plot the differences between the energies of $\eta_b$ states with momentum $(2,2,1)$ and $(3,0,0)$, in units of $2\pi/L$, on each ensemble using different actions. We see that the largest $SO(3)$ breaking occurs
for the $\order{p^4}$ NRQCD action. This breaking is reduced when including the $\order{\alpha_s p^4}$ kinetic couplings, and then reduced further by the $\order{p^6}$ NRQCD action. The least $SO(3)$ breaking occurs for the $\order{\alpha_s p^4, p^6}$ NRQCD action. This improvement is sizable for
the very coarse ensemble, Set 1, while for the coarse and fine
ensembles the improvement is visible, but small. Further, the breaking on the coarse and fine ensembles goes from being a significant effect to a non-significant ($2\sigma$) one after improvement. Using this
improved action allows for a more accurate and reliable determination
of the kinetic mass, and hence also of the tuned $b$-quark mass in high precision calculations. 

\begin{figure}[t]
  \centering
  \includegraphics[width=0.49\textwidth]{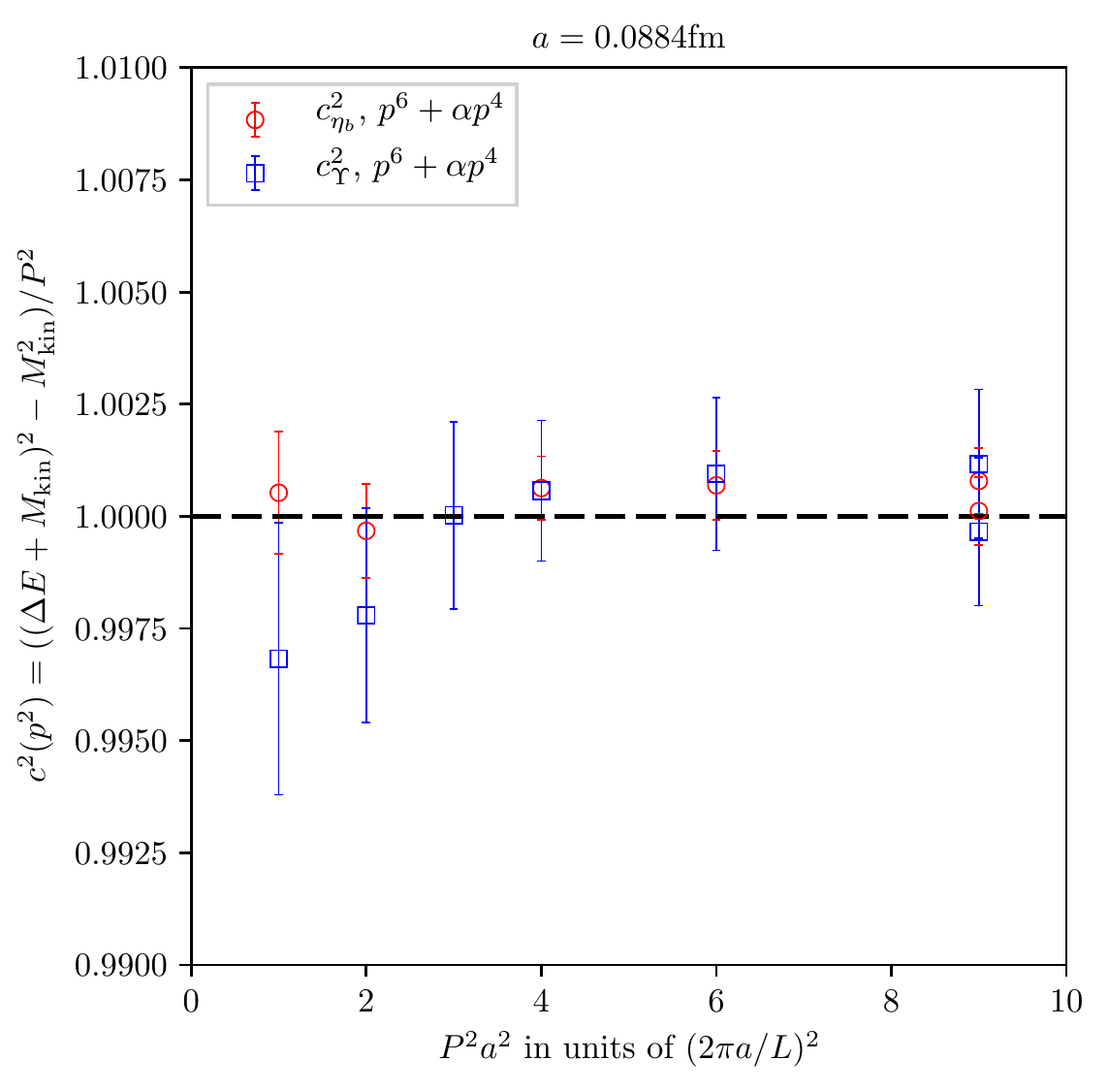}
  \caption{The speed of light squared, $c^2 P^2= (\Delta E + M_\text{kin})^2-M_\text{kin}^2$, on Set 5 with the $\order{\alpha_sp^4, p^6}$ NRQCD action. The errors shown are 
statistical only and we use the value of $M_\text{kin}$ computed using momentum $(1,1,1)$. The dashed line corresponds to $c^2 = 1$.}
  \label{fig:KinDeltaE}
\end{figure}

In Figure \ref{fig:KinDeltaE}, we show the speed of light squared, $c^2= \Big((\Delta E + M_\text{kin})^2-M_\text{kin}^2\Big)/P^2$, computed on Set 5 with the
$\order{\alpha_sp^4, p^6}$ NRQCD action against $P^2a^2$, where good agreement with the value of 1 is seen.

\section{Discussion and Conclusions}
\label{sec:Conclusions}

\begin{figure}[t]
  \centering
  \includegraphics[width=0.45\textwidth]{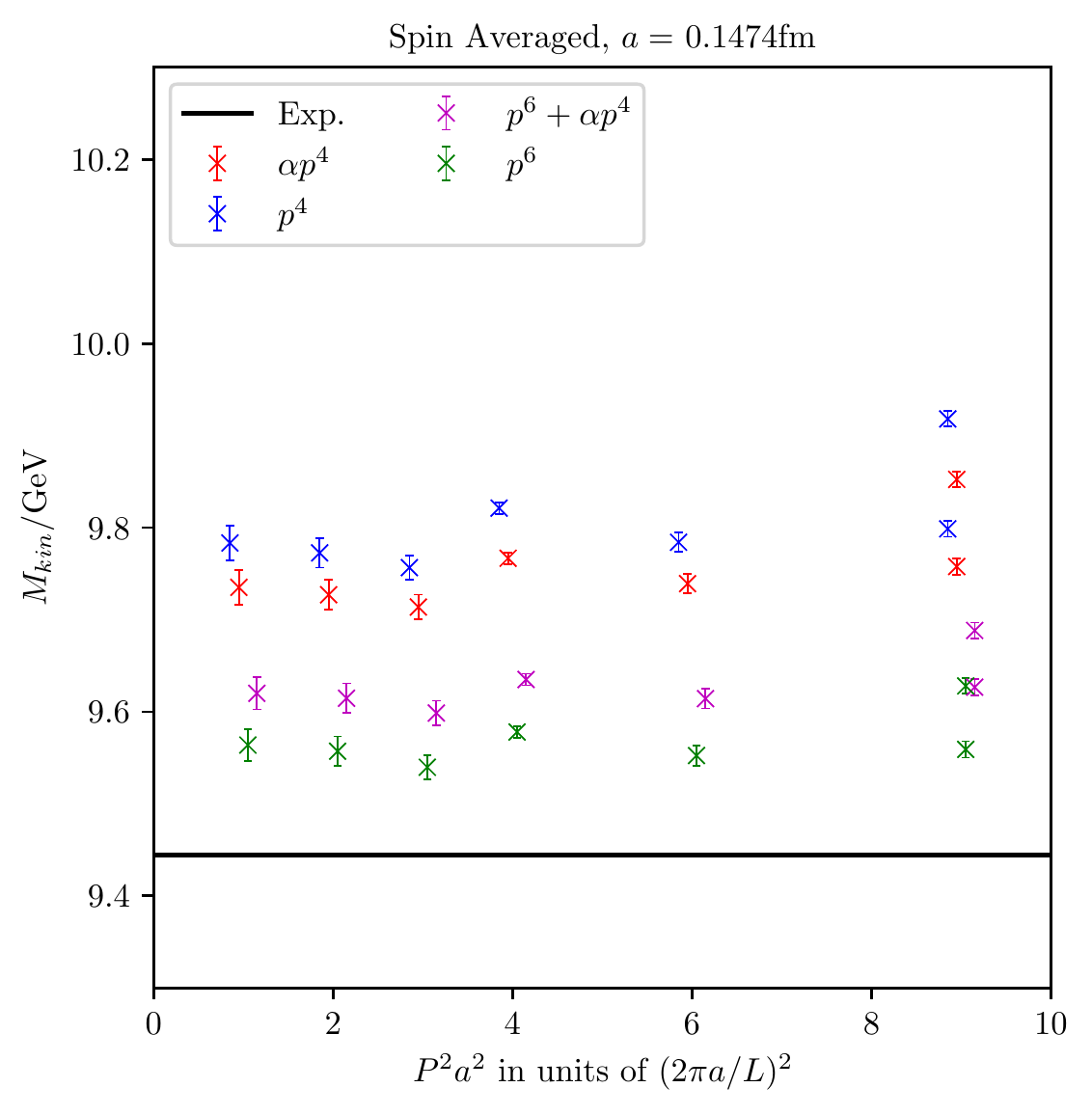} 
  \includegraphics[width=0.45\textwidth]{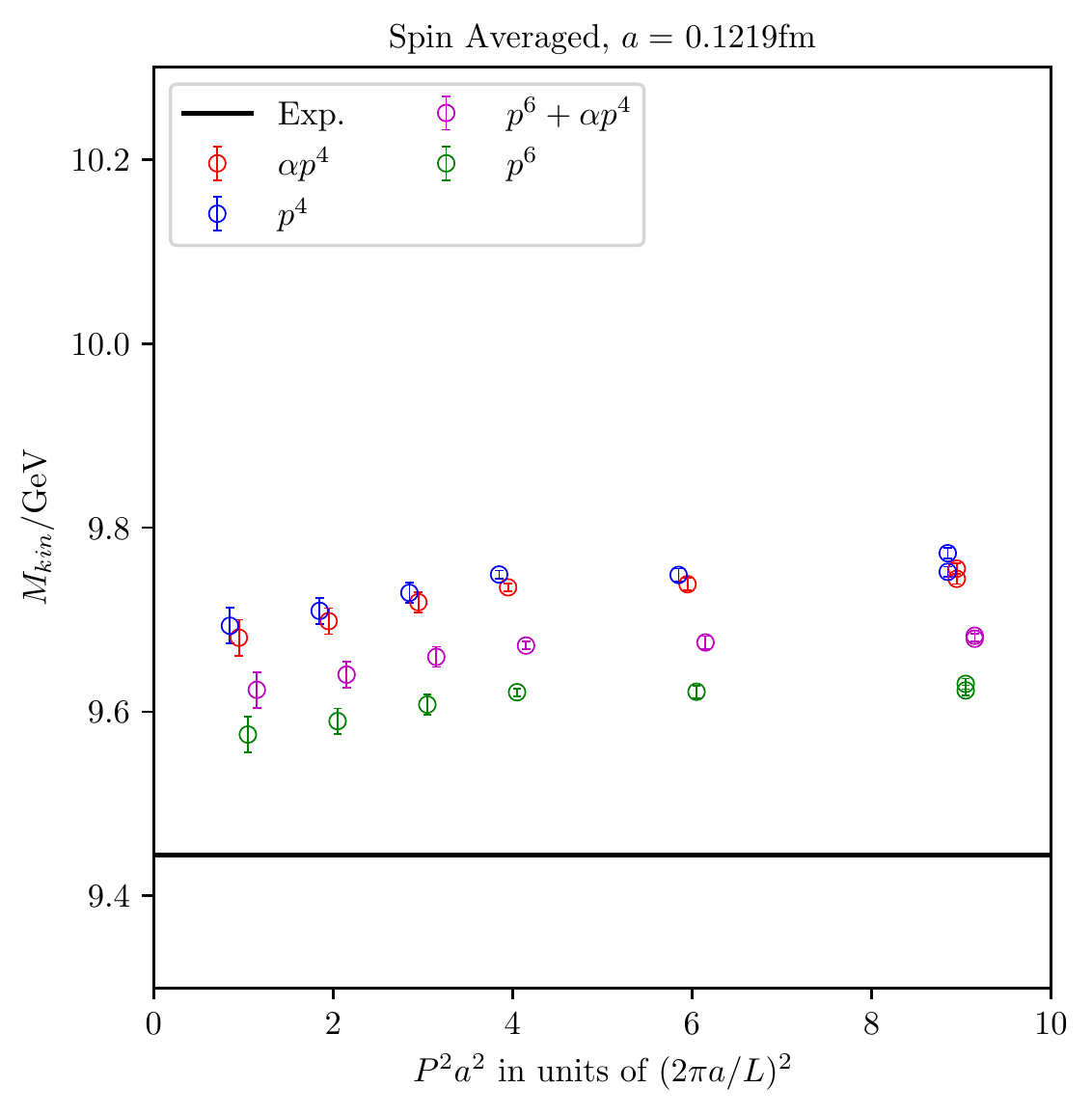}
  \caption{Non-perturbatively obtained spin-averaged kinetic masses, given by Eq. (\ref{spinaveq}), on very coarse Set 1 (above) and coarse Set 3 (below) for $p^4$ and $p^6$ actions with
    both tree-level and $\order{\alpha_s}$ $c_1,
    c_6$ and $c_5$. The errors shown are statistical only, excluding lattice spacing uncertainty, and are correlated. The data points at each value of $P^2a^2$ have been offset symmetrically for clarity. The larger energy with $a|{\VEC{P}}|=9$ is from the $(2,2,1)$ ground state. The solid line is the experimental value.  }
  \label{fig:KinE1}
\end{figure}

\begin{figure}[t]
  \centering
  \includegraphics[width=0.45\textwidth]{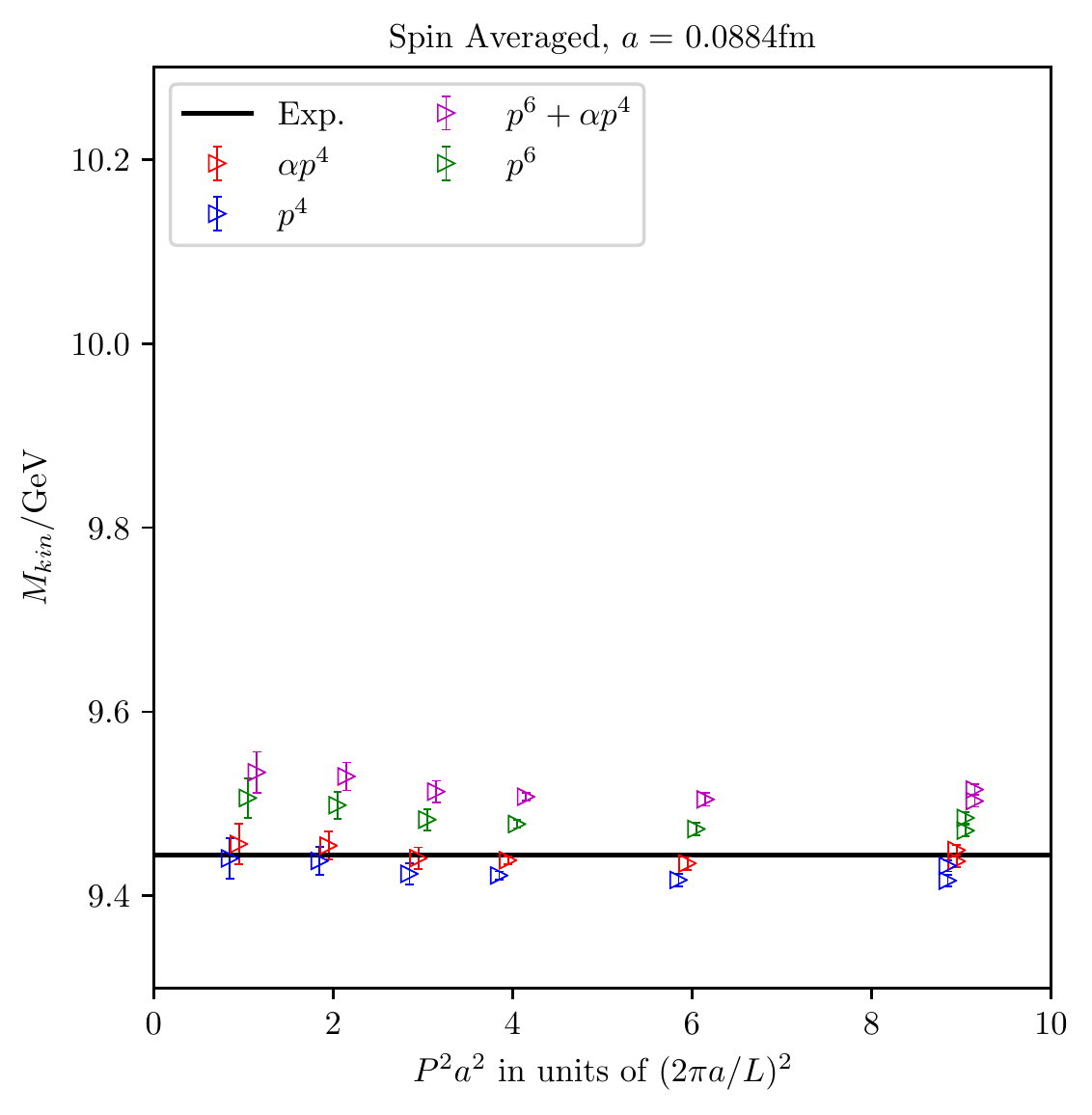}
  \caption{Same as Figure \ref{fig:KinE1} but on fine Set 5.}
  \label{fig:KinE3}
\end{figure}

\begin{figure}[t]
  \centering
  \includegraphics[width=0.45\textwidth]{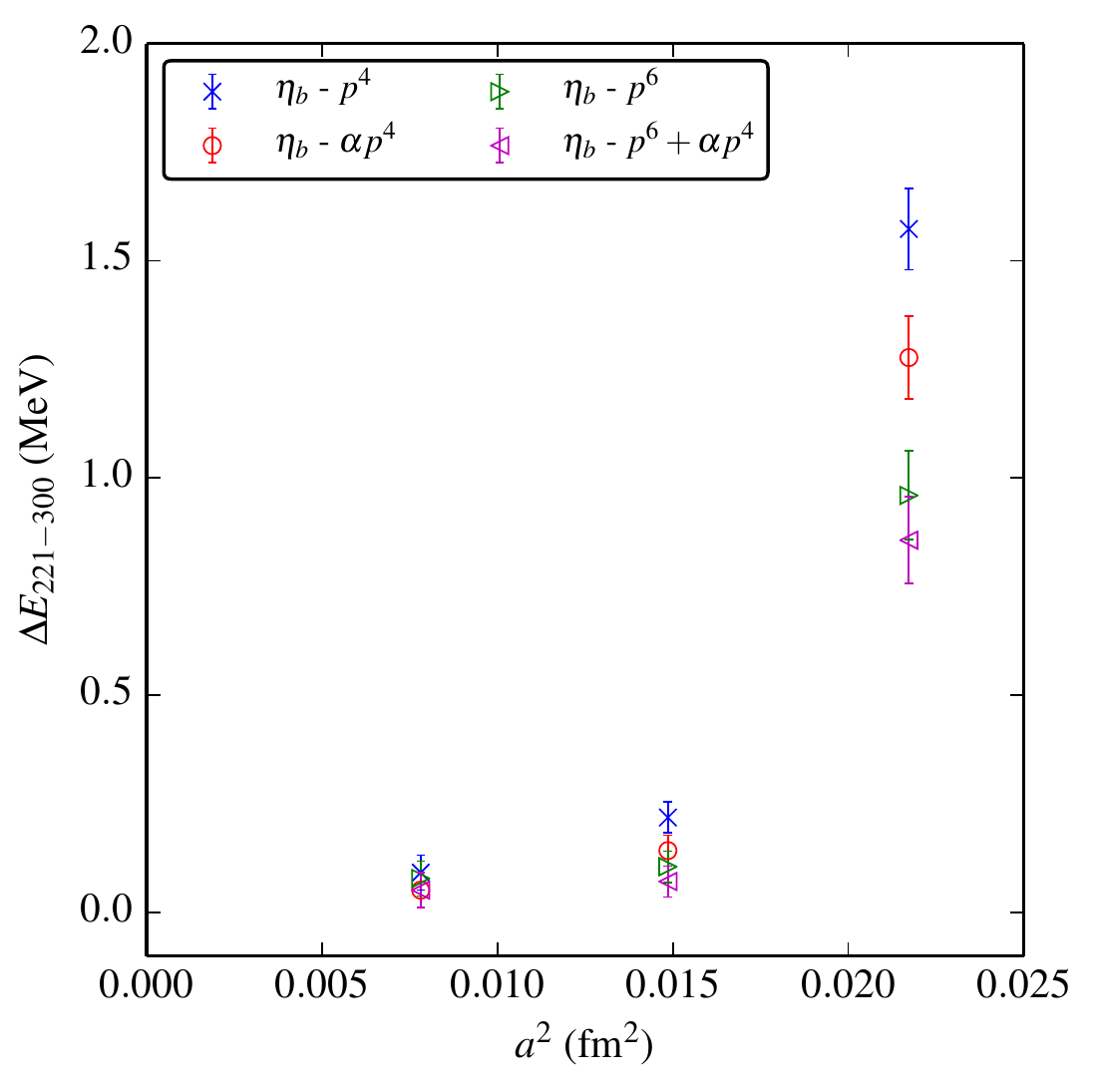}
  \caption{Non-perturbatively obtained values for $\Delta E_\mathrm{221-300}$ for the $\eta_b$ for each action plotted against the square of the lattice spacing.}
  \label{fig:deltaEetaups}
\end{figure}

In this work we have made the next
round of improvement to the HPQCD collaboration's formulation of the
NRQCD action to allow increasingly accurate non-perturbative calculations in the future. The key results presented herein include: 
\begin{itemize}
  \item Determining the required operators which need to be added to the NRQCD action
    in order to give a correct heavy-quark dispersion relation to
    $\order{p^6}$, presented in Sec. \ref{sec:NRQCDAction}.  
\item Determining the one-loop coefficients of the $\order{p^4}$
  kinetic couplings, namely $c^{(1)}_1, c^{(1)}_6$ and $c_5^{(1)}$, in automated lattice
  perturbation theory using twisted boundary conditions as an IR
  regulator. We also present results for the one-loop (bare-to-pole)
  heavy-quark mass renormalisation $Z_m^{(1)}$ and zero-point energy
  $W_0$ which can be combined to give the one-loop energy shift
  (from neglecting the quark mass term in the NRQCD action) of a
  $b$-quark. This one-loop energy shift can be added to the
  non-perturbatively determined simulation energies  to give a numerical value, which after
  converting to GeV, can be compared to the experimentally determined masses. All perturbative
  results are shown in Sec. \ref{sec:PertResults}.
\item Determining the full one-loop radiative correction of these
  quantities by finding the scale $q^*$ of $\alpha_s$ defined in the
  $V$-scheme. In doing this we use the higher order methodology which takes into account the anomalously small leading-order moments
  in order to obtain physical $q^*$ as described in Sec.~\ref{sec:Tads}. 
\item Determining the one-loop quantities for three different
  NRQCD action formulations, namely a NRQCD action that gives a
  heavy-quark dispersion relation correct (i) to $\order{p^4}$  and (ii) 
  to $\order{p^6}$. These actions employ a mean-field tadpole
  improvement procedure. For reasons described in Sec. \ref{sec:Tads} we also
  explore, for the first time, a (iii) Fat$3$ smeared NRQCD action
  with the quark dispersion   relation correct to $\order{p^4}$ which does not require mean-field improvement. The Fat$3$ results are encouraging and show stable behaviour against $am_b$, indicating that the use of this or a similar smearing may be the way forward in future, rather than tadpole-improvement.   
\item Varying the stability parameter with $n=4,6$ and $8$ to show
that, as shown in Sec.~\ref{sec:PertResults}, the $\order{\alpha_s p^4}$
kinetic couplings in the NRQCD action are 
  insensitive to this choice. Thus, if future calculations need to compensate a decrease in lattice spacing (which allows higher momentum fluctuations) with an increase in $n$, they can do so reliably. 
\item Testing how the improvement of the NRQCD action, both in terms
  of additional $\order{p^6}$ operators and one-loop radiative
  kinetic coefficients, affects the non-perturbatively obtained kinetic
  masses, c.~f.~Figs.~\ref{fig:KinE1}, \ref{fig:KinE3} and \ref{fig:deltaEetaups}. The impact of the $p^6$ terms and the radiative corrections on the kinetic masses obtained is small, particularly on the finer lattices. We find a significant reduction in $SO(3)$ symmetry breaking when using the improved actions on the very coarse ensemble, Set 1, which decreases as the lattice spacing is reduced. On the fine lattice, Set 5, $SO(3)$ symmetry breaking has been reduced to the point that the energy splitting, shown in Figure \ref{fig:deltaEetaups}, is very nearly consistent with zero. 
\end{itemize}

Taken together, NRQCD allows increasingly accurate and precise numerical
calculations to be performed by including higher-order operators, in combination with determining the matching coefficients using
perturbation theory. We have taken both these 
 steps in this work. Furthermore, NRQCD is numerically cheap compared to its relativistic
counterparts, being an initial-value, rather than a boundary-value
problem.  The outlook for NRQCD in the high-precision era is promising
and this work helps ensure that this NRQCD formalism will continue to
be an active contributor.

\section*{ACKNOWLEDGMENTS}

We would like to thank Christopher Thomas for the many insightful
discussions on finite-volume lattice effects. We are also grateful to the MILC collaboration for the use of their gauge configurations.
This manuscript has been authored by Fermi Research Alliance, LLC under Contract No.~DE-AC02-07CH11359 with the U.~S.~Department of Energy, Office of Science, Office of High Energy Physics. 
This work was funded in part by UK Science and  Technology Facilities Council (STFC) grants ST/L000385/1, ST/L000466/1 and ST/P000681/1. This work used the DiRAC Data Analytic system at the University of Cambridge, operated by the University of Cambridge High Performance Computing Service on behalf of the STFC DiRAC HPC Facility (www.dirac.ac.uk). This equipment was funded by BIS National E-infrastructure capital grant (ST/K001590/1), STFC capital grants ST/H008861/1 and ST/H00887X/1, and STFC DiRAC Operations grant ST/K00333X/1. DiRAC is part of the National E-Infrastructure.

\appendix

\section{Derivative Conventions}
\label{app:Der}

In this section we define our convention for the discretised derivative operators for use in the perturbative and non-perturbative calculations. The
forward, backward and  partially-cancelled second-order $\Delta^{(2)}_j$ operator are given by (note that all gauge-links are implicitly mean-field improved so that $U_i(x)$ is replaced by $U_i(x)/u_0$)
\begin{align}
&\Delta_i^+(x) = U_i(x)\psi(x + i) - \psi(x), \\
&\Delta_i^-(x) = \psi(x) - U_{-i}(x)\psi(x-i), \\
&\Delta^{(2),PC}_j\psi(x)   =  U_j(x) \psi(x+j)   \nonumber\\
& \hspace{3cm} +U_j^{\dagger}(x-j)\psi(x-j)- 2\psi(x). \label{eqn:D2} 
\end{align}
Then our partially corrected operators are 
\begin{align}
& \Delta^{(2)}    =  \sum_j \Delta^{(2),PC}_j, \\
& \Delta^{(4)}    =  \sum_j \Delta^{(2),PC}_j \Delta^{(2),PC}_j, \\
& \Delta^{(6)}    =  \sum_j \Delta^{(2),PC}_j \Delta^{(2),PC}_j \Delta^{(2),PC}_j, \\
& \Delta^{(2)} \Delta^{(4)}  = \left[\Delta^{(2)} \right] \left[ \Delta^{(4)} \right] +
\left(1 - \frac{1}{u_0^2} \right) ( \Delta^{(2)} - 18
), \label{eqn:D2D4PC} \\
& (\Delta^{(2)})^2  = \left[\Delta^{(2)}\right]^2 + 6 \left(1 -\frac{1}{u_0^2}\right), \label{eqn:D2D2} \\
& (\Delta^{(2)})^3   = \left[\Delta^{(2)}\right]^3   + \left(1 - \frac{1}{u_0^2} \right) \Big( 11 \Delta^{(2)} - 42 \Big). \label{eqn:D23PC}
\end{align} 

The additional terms in Eqs.~(\ref{eqn:D2D4PC}), (\ref{eqn:D2D2}) and
(\ref{eqn:D23PC}) are needed for the partial cancellation as described
in Sec. \ref{sec:Tads}. As can be seen, when the operators are
transformed to momentum space, the additional terms in these partially-cancelled operators
allow mixing down of higher-order coefficients to lower-order tadpole
counterterms. For the smeared operators, no mean-field improvement
is performed (i.e. $u_0$ is set to 1) so the additional terms vanish and
the links are replaced by their smeared counterparts. 

\section{Tadpole Counterterms from Mean-field Improvement}
\label{app:Tads}

In this appendix we will give explicit formulae for the tadpole
counterterms used to remove the unphysical tadpole contributions, as
described in Sec. \ref{sec:Tads}, when using a mean-field improvement
procedure. These formulae are utilised to produce the one-loop
mean-field improved quantities discussed in
Sec. \ref{sec:PertResults}. Features of these formulae have been
discussed in Sec. \ref{sec:Tads}.

\onecolumngrid
\begin{center}
\begin{table}[t]
  \caption{Numerical values of the one-loop coefficients. Note that the unsmeared results ($\mathcal{O}(p^4)$ and $\mathcal{O}(p^6)$) are mean-field improved given the formulae in
    Appendix \ref{app:Tads} and the mean field parameter in Table
    \ref{tab:Tads}. The smeared results (Fat3) are not mean-field improved.  \label{tab:coeffs}}
  \begin{tabular}{l l l l l l l}
    \hline \hline 
    $am_b$ &  $7.0$ & $4.0$ & $3.4$ & $2.8$ & $1.9$ & $1.1$ \\ \hline\hline
    $\tilde{c}_1^{(1)}$ \\ \hline 
    $\order{p^4}$ & $0.93051(24)$ & $0.68266(17)$ & $0.61998(16)$ &
    $0.55180(13)$ & $0.43443(15)$ & $0.261360(99)$ \\
    $\order{p^6}$ & $0.69607(24)$ & $0.52875(17)$ & $0.50348(16)$ &
    $0.48171(13)$ & $0.37427(10)$ & $-1.5525(50)$ \\ 
    Fat$3$        & $0.3991(37)$ & $0.2504(27)$ & $0.1914(24)$ &
    $0.1012(21)$ & $-0.1970(15)$ & $-1.25492(97)$ \\
    \hline\hline 
    ${c}_5^{(1)}$ \\ \hline 
    $\order{p^4}$ & $0.568(11)$ & $0.5341(62)$ & $0.5220(53)$ &
    $0.5056(44)$ & $0.4628(30)$ & $0.2946(17)$ \\ 
    $\order{p^6}$ & $0.07863(55)$ & $0.11875(32)$ & $0.13945(27)$ &
    $0.17215(22)$ & $0.26557(15)$ & $0.2212(58)$ \\ 
    Fat$3$        & $0.4813(87)$ & $0.4338(50)$ & $0.4100(42)$ &
    $0.3727(35)$ & $0.2569(24)$ & $-0.1199(69)$ \\
    \hline\hline 
    ${Z}_m^{(1)}$ \\ \hline 
    $\order{p^4}$ & $-0.08945(52)$ & $0.08892(52)$ & $0.15488(52)$ &
    $0.24259(52)$ & $0.44572(52)$ & $0.77983(52)$ \\ 
    $\order{p^6}$ & $-0.08988(52)$ & $0.08413(52)$ & $0.14947(52)$ &
    $0.24053(52)$ & $0.50827(52)$ & $1.85962(52)$ \\
    Fat$3$        & $0.08866(52)$ & $0.22842(52)$ & $0.28608(52)$ &
    $0.36957(52)$ & $0.60455(52)$ & $1.23018(52)$ \\
    \hline\hline 
    ${W }_0$ \\ \hline 
    $\order{p^4}$ & $-0.94595(52)$ & $-0.84252(52)$ & $-0.80232(52)$ &
    $-0.74890(52)$ & $-0.64040(52)$ & $-0.72316(52)$ \\
    $\order{p^6}$ & $-0.93235(52)$ & $-0.82391(52)$ & $-0.78118(52)$ &
    $-0.72050(52)$ & $-0.52587(52)$ & $1.16267(52)$  \\
    Fat$3$        & $-0.97946(52)$ & $-0.98997(52)$ & $-0.99508(52)$ &
    $-1.00270(52)$ & $-1.02215(52)$ & $-1.03670(52)$  \\
    \hline\hline 
    ${\delta C}$ \\ \hline 
    $\order{p^4}$ & $-0.22458(53)$ & $-0.12171(54)$ & $-0.08110(54)$ &
    $-0.02487(55)$ & $0.10867(59)$ & $0.12242(70)$  \\
    $\order{p^6}$ & $-0.22307(53)$ & $-0.12185(54)$ & $-0.08029(54)$ &
    $-0.01679(55)$ & $0.23149(59)$ & $2.91658(70)$ \\
    Fat$3$        & $-0.05126(53)$ & $-0.01907(54)$ & $-0.00659(54)$ &
    $0.01146(55)$ & $0.06658(59)$ & $0.30555(70)$ \\
    \hline\hline 
  \end{tabular}
\end{table}
\end{center}

\twocolumngrid

For a $\order{p^4}$ NRQCD action (e.g., using
Eq. (\ref{eqn:NRQCDGreensFunction}) with $\delta H_{p^6}=0$) with
partial cancellation, the tadpole counterterms are
\begin{align}
\frac{Z_m^{(1),tads}}{u_0^{(2)}} & = -\frac{2}{3} - \frac{3}{m_b^2}, \\
\frac{\tilde{c}_1^{(1),tads}}{u_0^{(2)}} & = -\frac{1}{8}\left(1 + \frac{m_b}{2n}
\right)^{-1} \Big[ \frac{12}{n^2} - \frac{1}{n}  \nonumber \\ 
& + \frac{1}{2m_b} \left(\frac{3}{n^2} - 4 \right) +
\frac{6}{m_b^2}\left( \frac{1}{n} -12 \right) + \frac{6}{m_b^3} \Big],
\nonumber \\
\frac{{c}_5^{(1),tads}}{u_0^{(2)}} & = -\frac{4}{3} + \frac{1}{4m_b} + \frac{3}{m_b^2} -
\frac{3}{8nm_b^2} - \frac{3}{4m_b^3}, \nonumber 
\end{align} 
\begin{align}
\frac{W_0^{tads}}{u_0^{(2)}} & = 1 + \frac{7}{2m_b} -
\frac{3}{2m_b^2} \left( 1 + \frac{m_b}{2n} \right). \nonumber 
\end{align}

For a $\order{p^6}$ NRQCD action (e.g., using
Eq. (\ref{eqn:NRQCDGreensFunction})) with
partial cancellation, the tadpole counterterms are
\begin{align}
& \frac{Z_m^{(1),tads}}{u_0^{(2)}}  = -\frac{3}{5} - \frac{43}{12m_b^2} + \frac{11}{4m_b^4}\left(1 - \frac{m_b^2}{6n^2}\right), \\
& \frac{W_0^{tads}}{u_0^{(2)}} = 1 + \frac{37}{10m_b} -
\frac{3}{4nm_b^2}-\frac{9}{4m_b^3} + \frac{21}{4m_b^5}\left( 1 -
\frac{m_b^2}{6n^2}  \right), \nonumber
\end{align}
\begin{widetext}
  \begin{align}
& \frac{\tilde{c}_1^{(1),tads}}{u_0^{(2)}}  = -\frac{1}{8}\left(1 +
  \frac{m_b}{2n} \right)^{-1} \bigg[ -\frac{56}{15} - \frac{7}{5n} +
\frac{1}{2m_b} \left(\frac{3}{n^2} - \frac{28}{5} \right)  +
\frac{1}{2m_b^2}\left( \frac{15}{n} - 28 \right) + \frac{9}{m_b^3} + \frac{66}{m_b^4}\left(1 - \frac{m_b^2}{6n^2}\right) \nonumber \\
&  \hspace{5.0cm} -\frac{21}{m_b^5}\left(1 - \frac{m_b^2}{6n^2}\right) - \frac{21}{2nm_b^4}\left(1 - \frac{m_b^2}{6n^2}\right)  \bigg], \nonumber \\
& \frac{{c}_5^{(1),tads}}{u_0^{(2)}} = -\frac{3}{5} + \frac{7}{20m_b} +
\frac{1}{m_b^2}\left(\frac{7}{12}-\frac{3}{8n} \right)  - \frac{9}{8m_b^3}
- \frac{11}{4m_b^4}\left(1 - \frac{m_b^2}{6n^2}\right) +
\frac{21}{8m_b^5}\left(1 - \frac{m_b^2}{6n^2}\right).  \nonumber \\
\end{align}
\end{widetext}

\section{Numerical Results}
\label{app:Data}

In this section, we give numerical values for the tadpole-improved
one-loop coefficients in Table \ref{tab:coeffs}. The tadpole
counterterms given in Appendix \ref{app:Tads} can be used with the
mean-field improved data to produce the raw results. We also give
$aq^*$ in Table \ref{tab:qstar} for each quantity which is used to determine the
the value of $\alpha_V$. Lastly, a subset of the full one-loop radiative
corrections relevant for heavy-quark non-perturbative calculations are
given in Tables \ref{tab:c1shift} and \ref{tab:c5shift}. Other values
can be read off the Figures in Sec. \ref{sec:PertResults}.

\onecolumngrid

\begin{center}
\begin{table}[t]
  \caption{Numerical values for the scale $aq^*$ used to evaluate $\alpha_s$ in the V-scheme when computing the one-loop contributions. Note that the unsmeared results ($\mathcal{O}(p^4)$ and $\mathcal{O}(p^6)$) are mean-field improved given the formulae in
    Appendix \ref{app:Tads} and the mean field parameter in Table
    \ref{tab:Tads}. The smeared results (Fat3) are not mean-field improved. The
    $aq^*$ are determined as described in Sec. \ref{sec:Scale}.    \label{tab:qstar}}
  \begin{tabular}{l l l l l l l}
    \hline \hline 
    $am_b$ &  $7.0$ & $4.0$ & $3.4$ & $2.8$ & $1.9$ & $1.1$ \\ \hline\hline
    $aq^*(\tilde{c}_1^{(1)})$ \\ \hline 
    $\order{p^4}$ & $2.4882(13)$ & $2.17692(51)$ & $2.06963(46)$ & $1.93677(66)$ & $1.6893(15)$ & $1.56045(59)$ \\
    $\order{p^6}$ & $2.6799(12)$ & $2.22771(80)$ & $2.10459(93)$ & $1.9845(15)$ & $1.7394(23)$ & $2.978(11)$ \\
    Fat$3$        & $2.459(22)$ & $2.083(20)$ & $1.961(21)$ & $1.737(27)$ & $2.056(14)$ & $1.7764(19)$ \\
    \hline\hline 
    $aq^*({c}_5^{(1)})$ \\ \hline 
    $\order{p^4}$ & $2.651(56)$ & $2.585(30)$ & $2.571(26)$ & $2.560(22)$ & $2.570(16)$ & $2.510(15)$ \\
    $\order{p^6}$ & $1.922(21)$ & $2.035(30)$ & $2.310(22)$ & $2.651(34)$ & $2.9942(87)$ & $0.625(19)$ \\
    Fat$3$        & $2.034(28)$ & $1.937(16)$ & $1.902(14)$ & $1.854(12)$ & $1.744(12)$ & $1.92(21)$ \\
    \hline\hline 
    $aq^*({Z}_m^{(1)})$ \\ \hline 
    $\order{p^4}$ & $1.375(35)$ & $1.169(11)$ & $1.0532(88)$ & $1.1508(62)$ & $1.1920(35)$ & $1.0931(36)$ \\ 
    $\order{p^6}$ & $1.370(21)$ & $1.170(12)$ & $0.9945(86)$ & $1.1342(61)$ & $1.3047(34)$ & $1.7829(25)$ \\
    Fat$3$        & $0.9598(80)$ & $1.0007(35)$ & $1.0416(29)$ & $1.0784(23)$ & $1.1316(24)$ & $1.2024(25)$ \\
    \hline\hline 
    $aq^*({W }_0)$ \\ \hline 
    $\order{p^4}$ & $1.0833(30)$ & $1.0370(32)$ & $1.0163(33)$ & $0.9874(34)$ & $0.9378(38)$ & $1.3173(47)$ \\
    $\order{p^6}$ & $1.0752(30)$ & $1.0248(32)$ & $1.0014(33)$ & $0.9636(35)$ & $0.9169(37)$ & $1.4451(84)$ \\
    Fat$3$        & $1.0118(27)$ & $1.0264(27)$ & $1.0337(27)$ & $1.0449(27)$ & $1.0761(27)$ & $1.1309(28)$ \\
    \hline\hline 
    $aq^*({\delta C})$ \\ \hline 
    $\order{p^4}$ & $1.4998(91)$ & $1.527(91)$ & $1.589(74)$ & $1.63(12)$ & $2.508(45)$ & $2.56(11)$ \\
    $\order{p^6}$ & $1.5868(48)$ & $1.63(12)$ & $1.69(11)$ & $1.79(11)$ & $2.428(21)$ & $2.3549(29)$ \\
    Fat$3$        & $1.866(26)$ & $1.390(76)$ & $0.7903(92)$ & $2.89(34)$ & $1.699(49)$ & $1.470(18)$ \\
    \hline\hline 
  \end{tabular}
\end{table}
\end{center}

\begin{center}
\begin{table}[t]
  \caption{ Numerical values of the one-loop radiative shift relevant for
    non-perturbative calculations. Note that the unsmeared results ($\mathcal{O}(p^4)$ and $\mathcal{O}(p^6)$) are mean-field improved given the formulae in
    Appendix \ref{app:Tads} and the mean field parameter in Table
    \ref{tab:Tads}. The smeared results (Fat3) are not mean-field improved. This
    data is plotted in Figure \ref{fig:sc1sc5}. To determine the physical scale,
$q^*$, we use $a^{-1} = 1.3$, $1.6$, $2.2$ and $3.3$ GeV corresponding to 
very coarse, coarse, fine and superfine MILC ensembles used by the
HPQCD collaboration \cite{Dowdall:Upsilon}. The $am_b$ values of 3.4, 2.8, 1.9 and 1.1 are the appropriate 
ones (approximately) for the $b$ quark on very coarse, coarse, 
fine and superfine ensembles respectively. However, here we 
give results for all 4 lattice masses at each lattice spacing 
for completeness.\label{tab:c1shift}}
  \begin{tabular}{l l l l l | l l l l } 
    \hline \hline 
    $\alpha_s(q^*)\tilde{c}_1^{(1)}$ \\ \hline     
    $am_b$ &  $3.4$ & $2.8$ & $1.9$ & $1.1$ &  $3.4$ & $2.8$ & $1.9$ & $1.1$\\ \hline\hline
    Very Coarse &   &       &       &      & Coarse & & &\\ \hline
    $\order{p^4}$ & $0.2138(39)$ & $0.1983(38)$ & $0.1713(37)$ & $0.1094(26)$ & $0.1905(29)$ & $0.1756(28)$ & $0.1494(27)$ & $0.0944(18)$ \\
    $\order{p^6}$ & $0.1719(31)$ & $0.1704(32)$ & $0.1445(31)$ & $-0.4417(64)$ & $0.1533(23)$ & $0.1513(24)$ & $0.1265(22)$ & $-0.4031(52)$  \\
    Fat$3$        & $0.0682(16)$ & $0.0391(12)$ & $-0.0682(14)$ & $-0.4775(99)$ & $0.0605(13)$ & $0.03424(97)$ & $-0.0607(11)$ & $-0.4190(73)$  \\
    \hline\hline 
    Fine &   &       &       &      & Superfine  &&&\\  \hline 
    $\order{p^4}$ & $0.1641(21)$ & $0.1503(20)$ & $0.1260(18)$ & $0.0788(12)$ & $0.1403(15)$ & $0.1278(14)$ & $0.1059(12)$ & $0.06573(79)$ \\
    $\order{p^6}$ & $0.1323(17)$ & $0.1298(17)$ & $0.1071(15)$ & $-0.3564(40)$ & $0.1132(12)$ & $0.1106(12)$ & $0.0902(10)$ & $-0.3117(31)$ \\
    Fat$3$        & $0.05184(97)$ & $0.02897(75)$ & $-0.05227(80)$ & $-0.3555(50)$ & $0.04413(75)$ & $0.02441(59)$ & $-0.04466(60)$ & $-0.3001(34)$ \\
    \hline\hline 
  \end{tabular}
\end{table}
\end{center}

\begin{center}
\begin{table}[t]
  \caption{The same as in Table \ref{tab:c1shift} but for $\alpha_s c_5^{(1)}$. 
    \label{tab:c5shift}}
  \begin{tabular}{l l l l l | l l l l } 
    \hline \hline 
    $\alpha_s(q^*){c}_5^{(1)}$ \\ \hline     
    $am_b$ &  $3.4$ & $2.8$ & $1.9$ & $1.1$ &  $3.4$ & $2.8$ & $1.9$ & $1.1$\\ \hline\hline
    Very Coarse &   &       &       &      & Coarse & & & \\ \hline
    $\order{p^4}$ & $0.1596(31)$ & $0.1549(28)$ & $0.1415(24)$ & $0.0912(15)$ & $0.1444(25)$ & $0.1402(23)$ & $0.1281(20)$ & $0.0824(13)$ \\
    $\order{p^6}$ & $0.04513(79)$ & $0.05182(85)$ & $0.0754(11)$ & \hspace{0.5cm} $-$  & $0.04055(62)$ & $0.04698(68)$ & $0.06879(86)$ & \hspace{0.5cm}$-$ \\
    Fat$3$        &  $0.1491(34)$ & $0.1377(31)$ & $0.0990(23)$ & $-0.0434(40)$ & $0.1318(26)$ & $0.1214(24)$ & $0.0867(18)$ & $-0.0384(32)$ \\
    \hline\hline 
    Fine &   &       &       &      & Superfine  &&&\\  \hline 
    $\order{p^4}$ & $0.1265(20)$ & $0.1228(18)$ & $0.1122(15)$ & $0.07210(95)$ & $0.1098(16)$ & $0.1065(14)$ & $0.0973(11)$ & $0.06244(72)$ \\
    $\order{p^6}$ & $0.03526(45)$ & $0.04125(51)$ & $0.06085(66)$ &
    \hspace{0.5cm} $-$ & $0.03037(33)$ & $0.03584(38)$ & $0.05324(49)$ & $0.0915(39)$ \\
    Fat$3$        & $0.1126(19)$ & $0.1035(17)$ & $0.0734(13)$ & $-0.0328(25)$ & $0.0956(15)$ & $0.0877(13)$ & $0.06187(92)$ & $-0.0279(20)$ \\
    \hline\hline 
  \end{tabular}
\end{table}
\end{center}

\twocolumngrid

\clearpage
\bibliographystyle{h-physrev5}
\bibliography{./CH}

\end{document}